\documentclass[fleqn,usenatbib]{mnras}

\usepackage[T1]{fontenc}
\usepackage{ae,aecompl}
\usepackage{graphicx}
\usepackage{amsmath}    
\usepackage{amssymb}
\usepackage{longtable}
\usepackage{verbatim}
\usepackage[singlelinecheck=off]{caption}
\usepackage{hyperref}
\usepackage{pdflscape}
\usepackage{threeparttable}
\usepackage{multirow}
\usepackage{lineno}
\usepackage{rotating}
\usepackage{eso-pic}

\AddToShipoutPictureBG*{%
  \AtPageUpperLeft{%
    \hspace{0.75\paperwidth}%
    \raisebox{-3.5\baselineskip}{%
      \makebox[0pt][l]{\textnormal{DES-2019-0438}}
}}}%

\AddToShipoutPictureBG*{%
  \AtPageUpperLeft{%
    \hspace{0.75\paperwidth}%
    \raisebox{-4.5\baselineskip}{%
      \makebox[0pt][l]{\textnormal{FERMILAB-PUB-19-621-AE}}
}}}%

\title[Ultracool dwarfs in wide binary and multiple systems]{Increasing the census of ultracool dwarfs in wide binary and multiple systems using Dark Energy Survey DR1 and Gaia DR2 data}

\author[DES Collaboration]{
\parbox{\textwidth}{
\Large
M.~dal Ponte,$^{1,2}$
B.~Santiago,$^{1,2}$
A.~Carnero~Rosell,$^{3,2}$
B.~Burningham,$^{4}$
B.~Yanny,$^{5}$
J.~L.~Marshall,$^{6}$
K.~Bechtol,$^{7,8}$ 
P.~Martini,$^{9,10}$
T.~S.~Li,$^{11,5,12,13}$
L.~De Paris,$^{1}$ 
T.~M.~C.~Abbott,$^{14}$
M.~Aguena,$^{15,2}$
S.~Allam,$^{5}$
S.~Avila,$^{16}$
E.~Bertin,$^{17,18}$
S.~Bhargava,$^{19}$
D.~Brooks,$^{20}$
E.~Buckley-Geer,$^{5}$
M.~Carrasco~Kind,$^{21,22}$
J.~Carretero,$^{23}$
L.~N.~da Costa,$^{2,24}$
J.~De~Vicente,$^{3}$
H.~T.~Diehl,$^{5}$
P.~Doel,$^{20}$
T.~F.~Eifler,$^{25,26}$
S.~Everett,$^{27}$
B.~Flaugher,$^{5}$ 
P.~Fosalba,$^{28,29}$
J.~Frieman,$^{5,12}$
J.~Garc\'ia-Bellido,$^{16}$
E.~Gaztanaga,$^{28,29}$
D.~W.~Gerdes,$^{30,31}$
D.~Gruen,$^{32,12,33}$
R.~A.~Gruendl,$^{21,22}$
J.~Gschwend,$^{2,24}$
G.~Gutierrez,$^{5}$
S.~R.~Hinton,$^{34}$
D.~L.~Hollowood,$^{27}$
K.~Honscheid,$^{9,35}$
D.~J.~James,$^{36}$
K.~Kuehn,$^{37,38}$
N.~Kuropatkin,$^{5}$
M.~A.~G.~Maia,$^{2,24}$
M.~March,$^{39}$
F.~Menanteau,$^{21,22}$
R.~Miquel,$^{40,23}$
A.~Palmese,$^{5,12}$
F.~Paz-Chinch\'{o}n,$^{21,22}$
A.~A.~Plazas,$^{11}$ 
E.~Sanchez,$^{3}$
V.~Scarpine,$^{5}$
S.~Serrano,$^{28,29}$
I.~Sevilla-Noarbe,$^{3}$ 
M.~Smith,$^{41}$
E.~Suchyta,$^{42}$ 
M.~E.~C.~Swanson,$^{22}$
G.~Tarle,$^{31}$
D.~Thomas,$^{43}$
T.~N.~Varga,$^{44,45}$
and A.~R.~Walker $^{14}$
\begin{center} (DES Collaboration) \end{center}
}
\\
\\
Affiliations are listed at the end of the paper.
}

\date{Accepted XXX. Received YYY; in original form ZZZ}

\pubyear{2018}
\begin{document}
\label{firstpage}

\maketitle

\begin{abstract}
We present the discovery of 255 binary and six multiple system candidates with wide (> 5$\arcsec$) separation composed by ultracool dwarfs companions to stars, plus nine double ultracool dwarf systems. These systems were selected based on common distance criteria. About 90\% of the total sample has proper motions available and 73\% of the systems also satisfy a common proper motion criterion. The sample of ultracool candidates was taken from the Dark Energy Survey (DES) and the candidate stellar primaries are from Gaia DR2 and DES data. We compute chance alignment probabilities in order to assess the physical nature of each pair. We find that 174 candidate pairs with Gaia DR2 primaries and 81 pairs with a DES star as a primary have chance alignment probabilities $< 5\%$. Only nine candidate systems composed of two ultracool dwarfs were identified. The sample of candidate multiple systems is made up of five triple systems and one quadruple system. The majority of the ultracool dwarfs found in binaries and multiples are of early L type and the typical wide binary fraction over the L spectral types is $2-4\%$. Our sample of candidate wide binaries with ultracool dwarfs as secondaries constitutes a substantial increase over the known number of such systems, which are very useful to constrain the formation and evolution of ultracool dwarfs.

\end{abstract}


\section{Introduction}

Ultracool dwarfs (UCDs) are presumed to be common objects in the Milky Way. But due to their very low masses, temperatures ($T_{\rm eff} < 2300K$) and hence luminosities, they are difficult sources to detect. Interest on them has increased in recent years. Very low mass stars have been found to harbour planetary systems, some of them similar to Earth \citep{Gillon2017}. Dust disks that could harbour protoplanetary systems have also been reported around young substellar sources (brown dwarfs) \citep{Sanchis2020}. And the incomplete census of such ultracool objects in the Galactic field, even close to the Sun, makes their initial mass function (IMF), spatial distribution and binary fraction relatively unconstrained and hard to place into the general context of Galactic star formation and evolution.

Large samples of M dwarfs, close to the H-burning limit already exist \citep{Lepine2011, West2011}. Also, the census of L and T dwarfs has greatly improved since the appearance of infrared surveys, such as the Two-Micron All-Sky Survey \citep[2MASS; ][]{Skrutskie2006}, the Deep Near Infrared Survey of the Southern Sky \citep[DENIS; ][]{Epchtein1997}, the UKIRT Infrared Deep Sky Survey \citep[UKIDSS; ][]{lawrence07}, the Wide-field Infrared Survey Explorer \citep[WISE;][]{Wright2010} and the VISTA Hemisphere Survey \citep[VHS; ][]{mcmahon13}. Among the optical surveys that unveiled substantial numbers of such ultracool sources are the Sloan Digital Sky Survey \citep[SDSS; ][]{York2000}, and, more recently, the Dark Energy Survey \citep[DES; ][]{Abb18} and Gaia DR2 \citep{Reyle2019}. 

On the theoretical side, uncertainties about the interiors, and, most specially, the atmospheres and evolution of L and T dwarfs still remain \citep{Pinfield2012, leggett2013, Baraffe2015}. As in the case of higher-mass stars, L and T dwarf formation and evolution models should benefit from knowledge of chemical composition, masses and ages of a sizeable sample of such objects. Binary systems are ideal for this purpose since the physical properties of the primary star can be applied to the UCD companion, assuming that the pair formed at the same time, of the same material and evolved in the same environment \citep{Faherty2010}. Also, large statistical samples could constrain intrinsic variations of the  formation process of the L and T dwarf population relative to more massive stars. 

In terms of binary statistics, there is evidence that the binary frequency decreases as a function of spectral type and separation. For solar-type stars, \citet{Raghavan2010} found that $\sim$ 25\% have a companion with separation wider than 100 astronomical units (AU), $\sim$ 11 \% wider than 1,000 AU and \citet{Tokovinin2012} estimate 4.4 \% wider than 2,000 AU. However, searches for M, L or T dwarfs in wide binary systems remain incomplete. Recently \citet{Dhital2011} and \citet{Dhital2015} presented the Sloan Low-mass Wide Pairs of Kinematically Equivalent Stars (SLoWPoKES), a catalogue containing common proper motion and common distance wide candidate pairs. For the mid-K and mid-M type dwarfs presented in both catalogues, the wide binary frequency was $\sim$ 1.1\%. The binary fraction for L and T dwarfs in wide systems is still uncertain. The fraction of L and T dwarfs found in binary and multiple systems, the distributions of mass ratios, primary spectral types, and separations may constrain different scenarios proposed for the formation of very low mass stars and brown dwarfs in general, and of binary and multiple systems involving these sources in particular \citep{Whitworth2004, Bate2005, Bonnell2008, Elmegreen2011, Reipurth2001, Jumper2013}. 

In this paper we present the search for wide binary and multiple systems which contain UCD companions, using the sample of 11,745 UCD candidates from \citet{Carnero2019}. We analyze the properties of this sample, including the distribution of projected separations and the binary fraction, and compare them to previous works.

In Section~\ref{sec:data} we describe the catalogues used and the criteria used to select the samples. In Section~\ref{sec:distance} we discuss the photometric distance measurement for the candidates and the spectrophotometric distance for the primary stars selected in the Gaia DR2 and DES data. In Section~\ref{sec:search} we present the the properties of candidate binaries and multiples and also we address the estimation of chance alignment probability. In Section~\ref{sec:discuss}, we show our analysis and comparisons to samples of wide binaries. Finally, we present our summary and concluding remarks in Section~\ref{sec:summary}.

\section{Candidate selection of ultracool dwarfs and primary stars}
\label{sec:data}

\subsection{DES, VHS and WISE data}
\label{subsec:des_vhs_wise}

DES is a ($\sim$5,000$deg^2$) optical survey in the $grizY$ bands used the Dark Energy Camera \citep[DECam;][]{flaugher2015}. DECam is a wide-field (3 $\deg^2$) imager at the prime focus of the Blanco 4m telescope in Cerro Tololo Inter-American Observatory (CTIO). 

The DES footprint was selected to obtain an overlap with the South Pole Telescope survey \citep{Carlstrom2011} and Stripe 82 from SDSS \citep{Abazajian2009}. The Galactic plane was avoided to minimize stellar foregrounds and extinction from interstellar dust in order to maintain the DES cosmological goals. Even though the main driver for DES is cosmological, the stellar data have been extensively used by the collaboration to identify new star clusters, streams and satellite galaxies in the MW Halo and beyond \citep{Bechtol2015, Drlica2015, Luque2017}.

The first public data release of the Dark Energy Survey, DES DR1 \citep[DR1;][]{Abb18} is composed of 345 distinct nights spread over the first 3 years of DES operations, from 2013 August 15 to 2016 February 12. The DES DR1 catalogue contains object flags including several that indicate corrupted values due to image artifacts or reduction problems. 

For the searches of UCDs, as discussed and presented in \citet{Carnero2019}, we demanded that \verb+FLAGS_z,Y+ = 0 (ensures no reduction problems in the $z$ and $Y$ bands) and \verb+ISO_MAGFLAGS_i,z,Y+ = 0 (ensures the object has not been affected by spurious events in the images in $i$,$z$,$Y$ bands). We also imposed a magnitude limit cut of $z <$ 22 with a detection of 5$\sigma$ at least in the $z$ and $Y$ to ensure a high completeness in the $i$ band, and therefore allow construction of colour-colour diagrams useful for the selection of UCDs. 

For the primary stars, we repeated this same approach, however we imposed a magnitude limit cut of $i <$ 24 and the quality cuts were performed in $g$,$r$,$i$ bands. The DES DR1 is a public release \footnote{\url{https://des.ncsa.illinois.edu/releases/dr1}}, but in this work we used \verb+SOF_PSF_MAG+ photometry, which has not been published yet. The SOF photometry is based on a different reduction using the \verb+ngmix+ code \footnote{\url{https://github.com/esheldon/ngmix}}, which has better PSF and shape modeling. Even though we used nonpublic photometry, the \verb+COADD_ID+ are the same as those in the public release.
 
In order to extend photometry into the infra-red, we matched the DES DR1 to the VHS and AllWISE data using a positional matching radius of $2\arcsec$. As discussed in \citet{Carnero2019}, for typical proper motions and a $2\arcsec$ match between DES and VHS, considering a 3-year baseline, our matching should yield a complete combined sample for distances $> 50 $ pc, with slowly decreasing completeness for more nearby and higher proper motion sources. After the match, we removed every source that did not pass the DES quality cuts as explained before. The resulting catalogues have 27,249,118 and 27,918,863 sources within a $2374 \ deg^{2}$ overlap region. These two catalogues were used for the UCD search (Section \ref{sec:LTcandidate}) and now to search for primary star candidates (Section \ref{sec:search_desvhs}), respectively.

\subsection{Sample of ultracool dwarf candidates}
\label{sec:LTcandidate}

As presented in \citet{Carnero2019}, our search of UCD candidates in the combination of DES, VHS and AllWISE data was performed using a colour-colour cut criteria. The adopted cuts to select our candidates was $(i_{AB}-z_{AB}) > 1.2$, $(z_{AB}-Y_{AB}) > 0.15$ and $(Y_{AB}-J_{Vega}) > 1.6$. We used this initial sample, mainly made up of M, L and T dwarfs, to run our spectral classification code, \textit{classif}, which uses only photometry, to estimate the spectral type of each object of the sample. The \textit{classif} code was implemented using the same method presented in ~\citet{skr14} and ~\citet{skr2016}, based on a minimization of the $\chi^2$ relative to M, L and T empirical templates. After running \textit{classif} we obtain 11,545 sources classified as L dwarfs and 200 as T dwarfs. More details about the selection method, colour cuts and the spectral classification can be found in \citet{Carnero2019}.

\subsection{Gaia DR2}
\label{subsec:gaia}

The Gaia astrometric mission was launched in December 2013. It is measuring positions, parallaxes, proper motions and photometry for over one billion sources to $G \simeq 20.7$. Its  Data Release 2 \citep[Gaia DR2;][]{gaiadr2}, has covered the initial 22 months of data taking (from a predicted total of 5 years), with positions and photometry for $1.7 \times 10^9$ sources and full astrometric solution for $1.3 \times 10^9$.

For our purpose, we used Gaia DR2 data to select primary star candidates. Particularly important for this work are the parallaxes, whose precision varies from $< 0.1$ mas for $G \leq 17$ to $\simeq 0.7$ mas for $G=20$. They allow us to better discern dwarf stars (whose distances will overlap those of the UCDs from DES, VHS and AllWISE) from much more distant giants of similar colours, $T_{\mathrm{eff}}$ and chemistry. For the stars brighter than G=18, the Gaia DR2 sample was cross-matched to the Pan-STARRS1 \citep{Kaiser2010}, 2MASS and AllWISE catalogues, so as to increase the amount of photometric information available for each star as we did for DES. The photo-astrometric distances, derived from precise parallaxes and photometry, are presented in \citet{Anders2019}. We refer to this sample as GaiaDR2-18.

\section{Distance and Proper Motion Measurements}
\label{sec:distance}

\subsection{Distance}

\subsubsection{Ultracool dwarf candidates}
\label{subsec:lt}

Using our UCD sample described in Section \ref{sec:LTcandidate}, we used the spectral type from each candidate and our empirical model grid described in \citet{Carnero2019} to estimate the absolute magnitude and then obtain the distance modulus for each UCD. 

The empirical model grid lists absolute magnitudes in $izYJHKW1W2$ for dwarfs ranging from M1 to T9. We computed one distance modulus for each filter with available apparent magnitude. The resulting distance to each UCD was then taken to be the mean value among the available filters and we used the dispersion around the mean as the distance uncertainty. We did not apply any correction for extinction, since this is expected to be small for the passbands we used and towards the relatively high Galactic latitudes covered by our samples. 

\subsubsection{Primary stars}
\label{subsec:gaia_des}

As mentioned before, we use the Gaia DR2~\citep{gaiadr2} and the combination of DES, VHS and AllWISE to search for stars located close to our UCD candidates. \cite{Anders2019} ran the \verb+StarHorse+ code ~\citep{Queiroz2018} on all stars in the Gaia DR2 sample brighter than $G = 18$, in an attempt to better constrain their distances and extinction, yielding what we call the GaiaDR2-18 sample. For DES stars, \verb+StarHorse+ was applied by us, but only to the stars that were close enough to the UCD candidates to be considered as a potential companion, as will be discussed in the next section. In this latter case, we use optical and infrared photometry, in addition to parallaxes from Gaia DR2 when available.

The \verb+StarHorse+ code uses a Bayesian approach to determine masses, ages, distances and extinctions for field stars through the comparison of their observed spectroscopic, photometric and astrometric parameters with those from stellar evolution models. The models used are the \verb+PARSEC+ set of isochrones \citep{Bressan2012}. The code assumes spatial priors for each structural component of the Galaxy (thin and thick disks, bulge and halo). The priors also assume Gaussian metallicity and age distribution functions for each structural component. For all components, the Chabrier Initial Mass Function \citep[IMF;][]{Chabrier2003} was assumed as a prior. Gaussian likelihood functions were generated using the available observed parameter set and their associated uncertainties. The code then computes the posterior distribution function over distance, marginalized for all other parameters. We take the median of this marginalized posterior as the best distance estimate, while the difference between the median 84th percentile and the (16th percentile) distances is taken as the higher (lower) 1-$\sigma$ uncertainty. For more details we refer to \citet{Queiroz2018} and \citet{Anders2019}.

\subsection{Proper motion}

The proper motion measurements for the primaries are mostly from Gaia DR2 catalog. However, for the UCDs, the proper motion measurements are from CatWISE Catalog \citep{Eisenhardt2019}. CatWISE is a catalog of selected sources from WISE and NEOWISE data collected from 2010 to 2016 in the W1 and W2 bands.

However, the majority of the UCDs distances are large and the motions are thus small compared to other samples. Also, the objects are faint and the time baselines relatively short, and so most of the proper motion uncertainties are comparable to the measurements themselves, making them consistent with zero. In this situation, proper motions may turn out to be an inefficient diagnostic of association. Nonetheless, we take into consideration these measurements in our binary and multiple systems search to assess their impact.

\section{The search for binary and multiple system candidates}
\label{sec:search}

Detection of faint sources close to brighter stars is difficult, with detections pushed to larger separations as the difference in brightness increases. We paired UCD candidates to potential primary stars using a search radius that corresponds to 10,000 AU as the projected separation between the pair members. Since the distances of our UCD candidates are in the 20 - 500 pc range, these search radii cover the angular range from 20$\arcsec$ to 500$\arcsec$. Details on how this projected separation is computed vary with the sample of primaries, as discussed in the next subsections. As discussed in \citet{marocco2017} and \citet{Deacon2014}, searches beyond 10,000 AU introduce a significant difficulty of disentangling widest binaries from chance alignments from field stars. 

To refine our wide binary and multiple systems, we checked if the members that have common distance also share a common proper motion, when available. The common distance criteria was made at the $2\sigma$ level. Also, the proper motions had to be within $2\sigma$ of each other. 

A robust binary or multiple system should satisfy
$\Delta_{\mu} \leq 2 \sigma_{\mu}$ where $\Delta_{\mu}$ is the total proper motion difference $$\Delta_{\mu} = \sqrt{\Delta_{\mu_{\alpha cos \delta}}^2 + \Delta_{\mu_{\delta}}^2}$$ and $\Delta_{\mu_{\alpha cos\delta}}$ and $\Delta_{\mu_{\delta}}$ are the differences in proper motion between the pair members in the two directions. In the above criterion, $$\sigma_{\mu} = \sqrt{\delta \mu_{1}^2 + \delta \mu_{2}^2}$$ is the composite uncertainty in the measured proper motions, where 1,2 represent the primary and secondary. The individual uncertainties in proper motion also combine in quadrature the uncertainties along each direction.

In the following sections, we describe how the pairing was done for each set, including the common distance and common proper motion requirements, and also discuss the way chance alignment probabilities were computed in each case.

\subsection{Ultracool dwarf companions to Gaia DR2 stars}
\label{sec:search_gaia}

For the GaiaDR2-18 primary candidate stars, we considered their \verb+StarHorse+ distances from \cite{Anders2019}, and used photometric distances to the UCD candidates. We defined a search radius equal to a projected separation of 10,000 AU evaluated at the lower limit in distance of the star, given its smaller distance uncertainty as compared to the UCD. For each star, we then searched for possible UCD companions within this projected radius. By additionally applying the common distance criterion, we found 174 candidate pairs. 

\begin{figure}
\begin{center}
    \includegraphics[width=\columnwidth]{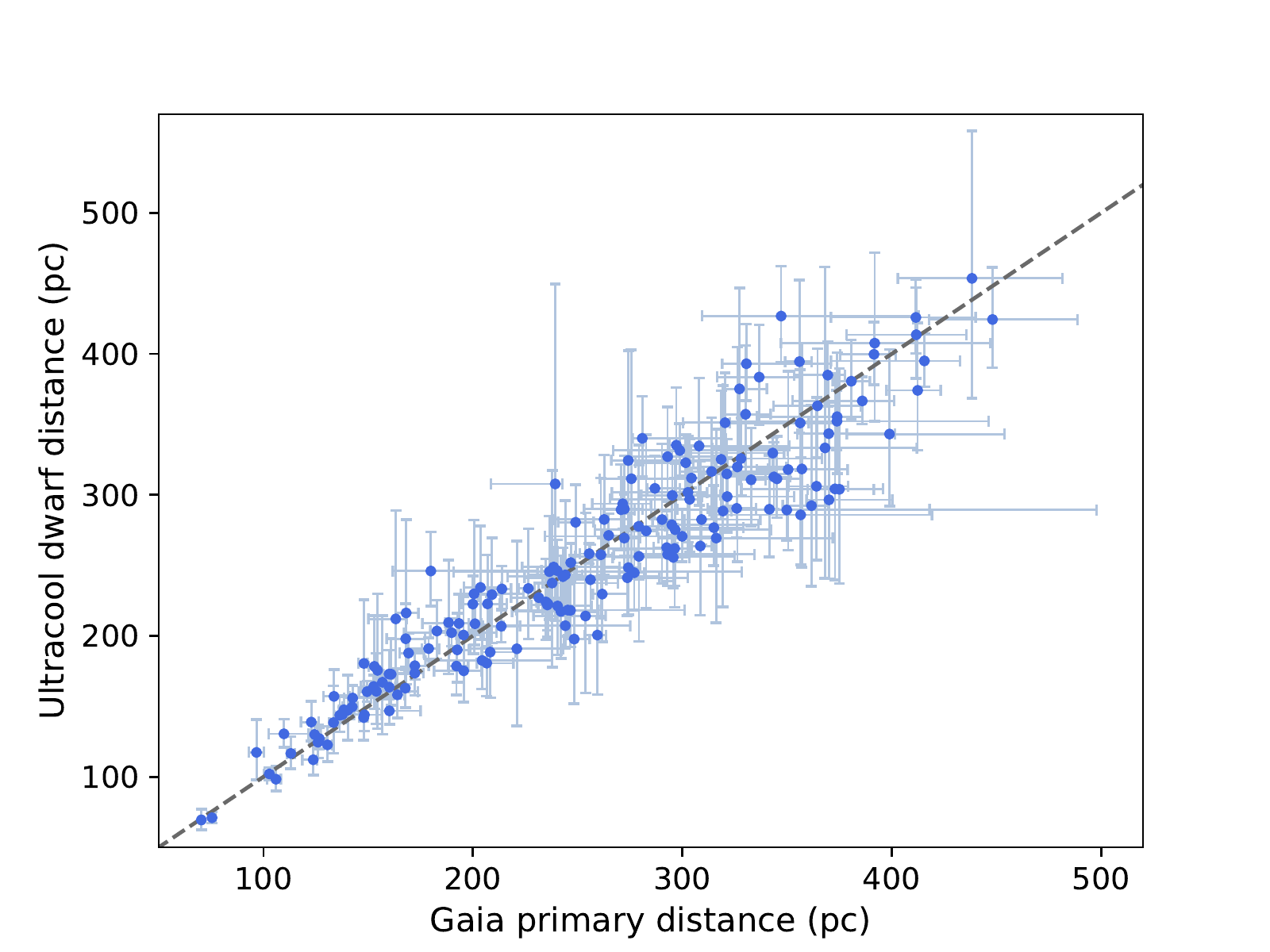}
\end{center}
\caption{The 174 common distance pair candidates identified using the UCD sample and Gaia DR2 primary candidate stars, taken from the sample by \citet{Anders2019}. The horizontal axis represents the primary distance given by StarHorse and the vertical axis shows the secondary's photometric distance. The error bars correspond to an uncertainty of 2$\sigma$. The uncertainties in the photometric distances of the UCD sample are usually much larger than those of the stars, which are based on measured parallaxes.}
\label{fig:ltgaia_pairs}
\end{figure}

For each possible pair, we estimate the chance alignment probability following a similar procedure used by \citet{smart2017} and \citet{Dhital2015}. The chance alignment probability is the probability that we find a physically unrelated object with the same common distance within our uncertainties and within the search radius. To assess the chance alignment probability, we simulate stars within a 2 $deg^2$ area from the UCD candidate from each pair using \verb+Trilegal+ \citep{Girardi2005}. The \verb+Trilegal+ simulated stars have a distance modulus without any uncertainty. In order to mimic an uncertainty in their distances, we use the uncertainty computed by \verb+StarHorse+ for the GaiaDR2-18 star whose distance is closest to that of the simulated \verb+Trilegal+ star. We thus assume that the uncertainty in distance for the simulated stars follows the same distribution as computed by \verb+StarHorse+ for real stars. We randomly selected 1,000 stars within the 2 $deg^2$ area and calculated the fraction $N/M$ of common distance stars, where N is the number of simulated stars which have the common distance with the UCD candidate and M is the total number of randomly selected simulated stars. Therefore, $N/M$ gives the probability of a randomly picked simulated star to have a common distance with the UCD. Then we obtain the probability over all stars within the search radius by multiplying $N/M$ by the number of simulated stars and making an area normalization considering the search radius area and the simulated area. We flag every pair with a chance alignment probability $P_{a}$ > 5\% as contamination.

In the current sample based on GaiaDR2-18 primaries, all of the 174 common distance pairs survived the $P_{a}$ < 5\% cut. These candidate wide binaries are shown in Figure \ref{fig:ltgaia_pairs}. A simple estimate of the number of chance alignments that still made into the sample may be obtained by adding up the $P_a$ values, yielding a total of 1.078. Among the 174 candidate pairs, 153 UCDs had proper motion in CatWISE catalog. Applying the common proper motion criteria, a sample of 125 pairs remains. This shows that 82\% of the common distance systems survive the proper motion refinement criterion, at the expense of losing a fraction of the objects due to lack of proper motion data. The properties for a subset of these candidate pairs are presented in Table \ref{tab:gaia_lt}. The entire table is available in machine-readable format in \url{https://des.ncsa.illinois.edu/releases/other/y3-lt-widebinaries}.

\begin{figure}
\begin{center}
    \includegraphics[width=\linewidth]{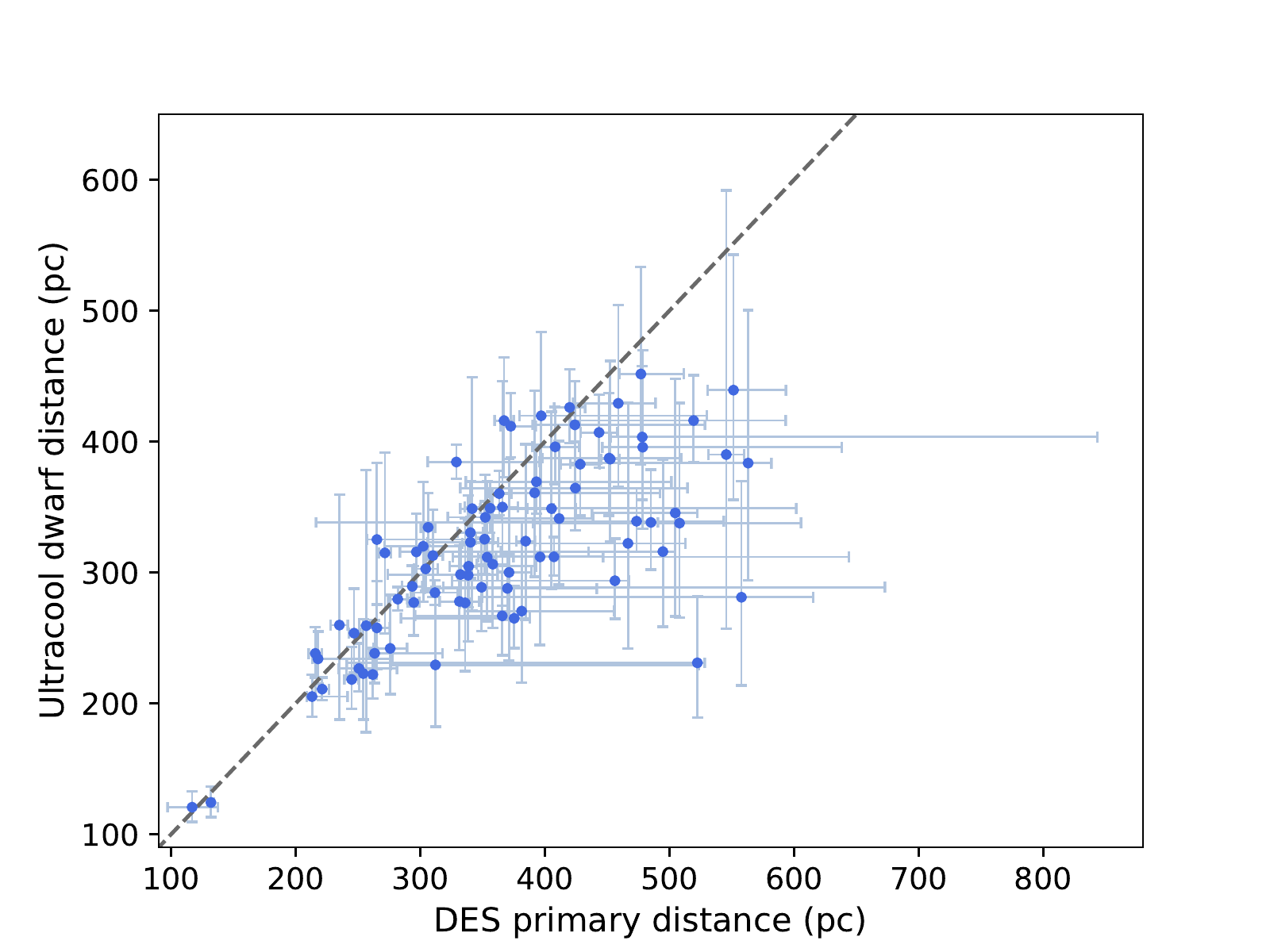}
\end{center}
\caption{The 85 common distance pair candidates identified using the UCD sample and DES primary stars. The horizontal axis represents the primary photometric distance given by StarHorse and the vertical axis shows the UCD photometric distance. The error bars indicates an uncertainty of 2$\sigma$.}
\label{fig:ltdes_pairs}
\end{figure}

\subsection{Ultracool companions to DES DR1 stars}
\label{sec:search_desvhs}

In this case, the search radius corresponds to 10,000 AU projected separation evaluated at the lower distance limit for the UCD. We adopt this threshold because we do not have the \verb+StarHorse+ distances for the entire DES stars catalogue. Due to computational restrictions, we only obtain the \verb+StarHorse+ distance for stars that were inside the UCD search radius. Considering that these UCDs have a large uncertainty in their purely photometric distances, this conservative approach should result in a larger search radius and the inclusion of several stars within this radius.

As mentioned in the previous section, in this case \verb+StarHorse+ distances for the primary stars were based on photometric measurements, with additional constraint from parallaxes for a small number of DES primary which are common to Gaia DR2. We thus applied the common distance criterion and were able to find 85 possible pairs involving a DES DR1 primary and a UCD as a secondary, as shown the Figure \ref{fig:ltdes_pairs}. 

As we explain in the previous section, for the chance alignment probabilities, we rely on \verb+Trilegal+ simulations. The procedure is the same as described in Section \ref{sec:search_gaia}. We assign distance uncertainties to the simulated stars using the closest DES DR1 star. For each secondary, we randomly selected 5,000 stars in the simulated area and require that the distances of the UCD candidate and the simulated star lie within 2$\sigma$ of each other. Thus, we obtain the probability over all simulated stars within the search radius. In the case of the 85 candidate wide binaries identified with DES DR1, 81 of them have $P_{a}$ < 5\%. The sum of the $P_a$ values for this sample yields 1.468 as the expected number of remaining unphysical pairs.

From the 81 candidate pairs, 74 UCDs have proper motion measurements from CatWISE. After applying the common proper motion criteria, 61 pairs remained in the sample, again yielding a fraction of 82\% pairs that pass the cut in proper motions. The properties for a subset of these candidate pairs are also presented in Table \ref{tab:gaia_lt}. 

\begin{figure}
\begin{center}
    \includegraphics[width=\linewidth]{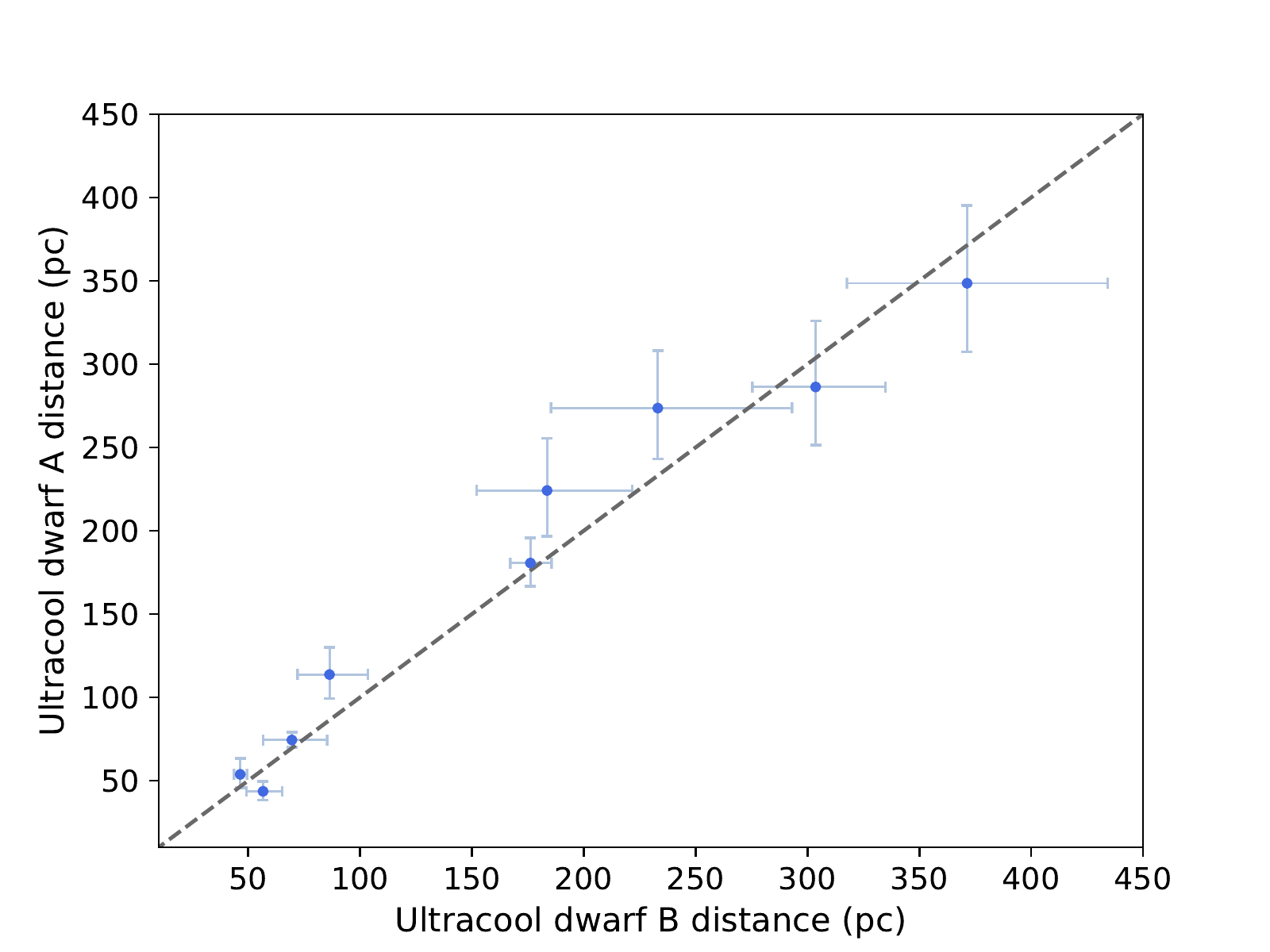}
\end{center}
\caption{The nine common distances for the pure UCD binary candidates identified. The horizontal and vertical axis show the UCDs photometric distances and the error bars correspond to an uncertainty of 2$\sigma$.}
\label{fig:ltlt_pairs}
\end{figure}

\subsection{Wide binaries involving two ultracool dwarfs}
\label{subsec:search_desLTs}

We also used the UCD sample to search for candidate binaries among themselves. We computed a search radius for each UCD and checked if another such dwarf appears inside this individual radius. We were able to identify nine possible pairs, which are shown in Figure \ref{fig:ltlt_pairs}. The properties of these possible binary pairs are presented in Table \ref{tab:ltlt}. The pairs are matched independently of the pair member that we centered on, except for one system. In other words, if source B is found within the search radius of 10,000 AU around source A, this latter was also within the same projected separation at B's distance. The entire table is available in \url{https://des.ncsa.illinois.edu/releases/other/y3-lt-widebinaries}.

\begin{landscape}
\thispagestyle{empty}

\begin{table}
\caption{The common distance pair candidates identified using the UCD sample combined to Gaia DR2 and DES DR1 data. The ID in $Jhhmm \pm ddmm$ format is based on the primary coordinates. The ID with a * symbol indicates that the distance for Gaia DR2 stars was published by \citet{Anders2019} and the $\dag$ symbol indicates common proper motion systems. The letter A represents the primary star and B the UCD. The angular and the projected separation are indicated by $\Delta\theta$ and $d_{P}$, respectively. The $P_{a}$ refers to the chance alignment probability, as explained in Section \ref{sec:search_gaia} and \ref{sec:search_desvhs}. The entire table is available in \url{https://des.ncsa.illinois.edu/releases/other/y3-lt-widebinaries}.}
\label{tab:gaia_lt}
%\begin{tabular}{\textwidth}{@{\extracolsep{\fill}}lcccccccccccccc}
%\begin{threeparttable}
\begin{tabular}{lccccccccccccccc}
\hline\hline
\multicolumn{1}{c}{ID} & \multicolumn{4}{c}{Position} & \multicolumn{4}{c}{Photometry} & \multicolumn{2}{c}{Distance} & Sp. Type & \multicolumn{3}{c}{Binary Information} \\
& $\alpha_{A}$ & $\delta_{A}$ & $\alpha_{B}$ & $\delta_{B}$ & $G_{Gaia, A}$ & $z_{DES,A}$ & $i_{DES,B}$ & $z_{DES,B}$ & $d_{A}$ & $d_{B}$ & B & $\Delta\theta$ ($\arcsec$) & $d_{P}$ (AU) & $P_{a}$ (\%)\\
\hline
J0001$-$4315*$\dag$ & 00:01:52 & -43:15:45 & 00:01:51 & -43:15:41 & 17.2 & 15.9 & 23.0 & 21.4 & 295$\pm$18  & 279$\pm$8  & L1 & 15.0 & 4427.4  & 0.264\\
J0002+0006*$\dag$   & 00:02:10 & +00:06:28 & 00:02:08 & +00:07:06 & 15.4 & 14.1 & 22.0 & 20.4 & 172$\pm$3   & 179$\pm$5  & L1 & 43.1 & 7440.3  & 0.222 \\
J0002$-$0626*$\dag$ & 00:02:24 & -06:26:11 & 00:02:23 & -06:26:30 & 16.7 & 15.2 & 20.6 & 19.1 & 130$\pm$2   & 123$\pm$6  & L0 & 34.3 & 4482.0  & 0.391  \\
J0003$-$5803*  & 00:03:24 & -58:03:51 & 00:03:18 & -58:04:06 & 14.1 & 12.8 & 21.2 & 19.9 & 154$\pm$1   & 175$\pm$24  & L0 & 50.3 & 7776.1  & 1.389  \\
J0005+0104$\dag$    & 00:05:46 & +01:04:54 & 00:05:47 & +01:04:43 & 18.8 & 17.2 & 23.0 & 21.5 & 393$\pm$62  & 369$\pm$13  & L0 & 14.1 & 5580.9  & 1.137  \\
J0008$-$4929*  & 00:08:07 & -49:29:27 & 00:08:08 & -49:29:20 & 17.7 & 16.4 & 22.6 & 21.3 & 314$\pm$8  & 316$\pm$18  & L0 & 16.0 & 5054.3  & 0.649  \\
J0008$-$0437   & 00:08:15 & -04:37:53 & 00:08:17 & -04:37:38 & 18.1 & 16.7 & 22.4 & 20.8 & 375$\pm$30  & 265$\pm$12  & L0 & 26.6 & 10017.4 & 1.100 \\

\hline
\end{tabular}

%\end{threeparttable}
\end{table}

\begin{table}
\caption{The nine common distance pair candidates identified among the UCD sample and the letters A and B represent a different UCD. The ID in $Jhhmm \pm ddmm$ format is based on the primary coordinates. The $\dag$ symbol indicates indicates common proper motion systems. The angular and the projected separation are indicated by $\Delta\theta$ and $d_{P}$, respectively. The $P_{a}$ refers to the chance alignment probability.}
\label{tab:ltlt}
\begin{tabular}{lccccccccccccccc}
%\begin{tabular*}{\textwidth}{@{\extracolsep{\fill}}cccccccccccccc}
\hline\hline
\multicolumn{1}{c}{ID} & \multicolumn{4}{c}{Position} & \multicolumn{4}{c}{Photometry} & \multicolumn{2}{c}{Distance} & \multicolumn{2}{c}{Sp. Type} & \multicolumn{3}{c}{Binary Information} \\
& $\alpha_{A}$ & $\delta_{A}$ & $\alpha_{B}$ & $\delta_{B}$ & $i_{DES,A}$ & $z_{DES,A}$ & $i_{DES,B}$& $z_{DES,B}$ & $d_{A}$ & $d_{B}$ & A & B & $\Delta\theta$ ($\arcsec$) & $d_{P}$ (AU) & $P_{a}$ (\%) \\
\hline
J0003$-$0011 & 00:03:30 & -00:11:06 & 00:03:35 & -00:12:59 & 23.1 & 21.0 & 20.6 & 19.1 & 70 $\pm$7 & 74 $\pm$2  & L7 & L2 & 131.5 & 9141.83 & 0.078 \\
J0443$-$4551$\dag$ & 04:43:10 & -45:51:55 & 04:43:04 & -45:50:23 & 25.0 & 21.9 &  -   & 21.7 & 57 $\pm$4  & 44 $\pm$3  & T5 & T6 & 112.0 & 6345.85 & 0.020  \\
J0457$-$4933$\dag$ & 04:57:49 & -49:33:56 & 04:57:52 & -49:34:02 & 22.6 & 21.1 & 22.4 & 20.9 & 304$\pm$15 & 286$\pm$19 & L0 & L0 & 25.61 & 7777.43 & 0.007 \\
J2000$-$5342$\dag$ & 20:00:12 & -53:42:38 & 20:00:12 & -53:43:07 & 23.1 & 21.3 & 22.3 & 20.9 & 233$\pm$27 & 274$\pm$16 & L2 & L0 & 29.41 & 6857.05 & 0.115 \\
J2251$-$4959$\dag$ & 22:51:57 & -49:59:32 & 22:51:56 & -50:00:01 & 22.9 & 21.4 & 23.0 & 21.4 & 371$\pm$29 & 349$\pm$22 & L0 & L0 & 29.51 & 10961.6 & 0.170 \\
J2313$-$4550$\dag$ & 23:13:49 & -45:50:29 & 23:13:49 & -45:50:25 & 22.2 & 20.3 & 23.3 & 21.2 & 86 $\pm$8 & 114$\pm$8 & L4 & L5 & 4.080 & 352.297 & 0.000 \\
J2318$-$5420 & 23:18:39 & -54:20:34 & 23:19:03 & -54:21:49 & 18.4 & 17.0 & 21.4 & 19.9 & 46 $\pm$1  & 54 $\pm$4  & L0 & L5 & 226.4 & 10514.6 & 0.025  \\
J2319$-$5203$\dag$ & 23:19:43 & -52:03:55 & 23:19:48 & -52:04:24 & 21.4 & 19.9 & 21.4 & 19.9 & 176$\pm$5 & 181$\pm$7 & L0 & L0 & 59.10 & 10411.8 & 0.045\\
J2319$-$5607$\dag$ & 23:19:50 & -56:07:27 & 23:19:47 & -56:07:56 & 22.1 & 20.6 & 22.4 & 21.0 & 184$\pm$17 & 224$\pm$15 & L1 & L1 & 37.21 & 6832.95 & 0.066 \\
\hline
\end{tabular}
\end{table}

\begin{table}
\caption{The common distance multiple systems found in our search. The letters A and B represent a star from Gaia DR2 or DES DR1 and the letter C the UCD. In the last row, B and C represent a different UCD. The ID in $Jhhmm \pm ddmm$ format using the primary coordinates. The ID with a * symbol indicates that the distance for Gaia DR2 stars was published by \citet{Anders2019} and the $\dag$ symbol indicates common proper motion systems.} 
\label{tab:multi_sys}
\begin{tabular}{lcccccccccccccc}
\hline \hline 
\multicolumn{1}{c}{ID} & \multicolumn{6}{c}{Position} & \multicolumn{3}{c}{Photometry} & \multicolumn{3}{c}{Distance} & \multicolumn{2}{c}{Sp. Type}  \\
& $\alpha_{A}$ & $\delta_{A}$ & $\alpha_{B}$ & $\delta_{B}$ & $\alpha_{C}$ & $\delta_{C}$ & $G_{Gaia,A}$ & $G_{Gaia,B}$ & $z_{DES,C}$ & $d_{A}$ & $d_{B}$ & $d_{C}$ & B & C \\
\hline

J0009$-$5313*$\dag$ & 00:09:06 & -53:13:30 & 00:09:06 & -53:13:34 & 00:09:03 & -53:13:43 & 16.8 & 17.0 & 23.0 & 241$\pm$3 & 231$\pm$3 & 235$\pm$17 & & L2  \\
J0042$-$0331*$\dag$ & 00:42:33 & -03:31:29 & 00:42:33 & -03:31:30 & 00:42:34 & -03:31:53 & 17.9 & 16.4 & 21.7 & 200$\pm$10 & 193$\pm$10 & 195$\pm$16 & & L0 \\
J0239$-$0512* & 02:39:43 & -05:12:59 & 02:39:44 & -05:12:54 & 02:39:45 & -05:13:17 & 17.0 & 20.6 & 20.9 & 356$\pm$11 & 345$\pm$3 & 286$\pm$15 & & L0 \\
J2024$-$5801*$\dag$ & 20:24:13 & -58:01:15 & 20:24:14 & -58:01:15 & 20:24:16 & -58:01:15 & 16.9 & 16.2 & 22.3 & 232$\pm$6 & 222$\pm$4 & 218$\pm$7 & & L1 \\
J2200$-$4155*$\dag$ & 22:00:25 & -41:55:02 & 22:00:25 & -41:55:02 & 22:00:25 & -41:55:37 & 17.3 & 17.3 & 20.9 & 198$\pm$7 & 202$\pm$8 & 214$\pm$3 & & L1 \\
J2342$-$6135 & 23:42:06 & -61:35:44 & 23:42:08 & -61:35:42 & 23:42:04 & -61:35:17 & 20.2 & - & 20.7 & 267$\pm$28 & 286$\pm$29 & 279$\pm$29 & L0 & L0 \\

\hline
\end{tabular}

\end{table}

\end{landscape}

To obtain the chance alignment probability we used the \verb+GalmodBD+ simulation code, presented in \cite{Carnero2019}, which computes expected Galactic counts of UCDs as a function of magnitude, colour and direction on the sky. \verb+GalmodBD+ also creates synthetic samples of ultracool dwarfs based on the expected number counts for a given footprint, using empirically determined space densities of objects, absolute magnitudes and colours as a function of spectral type. For the current purpose, we computed the expected number of UCDs in a given direction and within the volume bracketed by the common range of distances and by the area within the angular separation of each possible pair. 
For all the nine candidate pairs, the probability of chance alignment is $P_{a}$ < 0.2 \%, as shown in Table \ref{tab:ltlt}. 

\begin{figure}
\begin{center}
    \includegraphics[width=\linewidth]{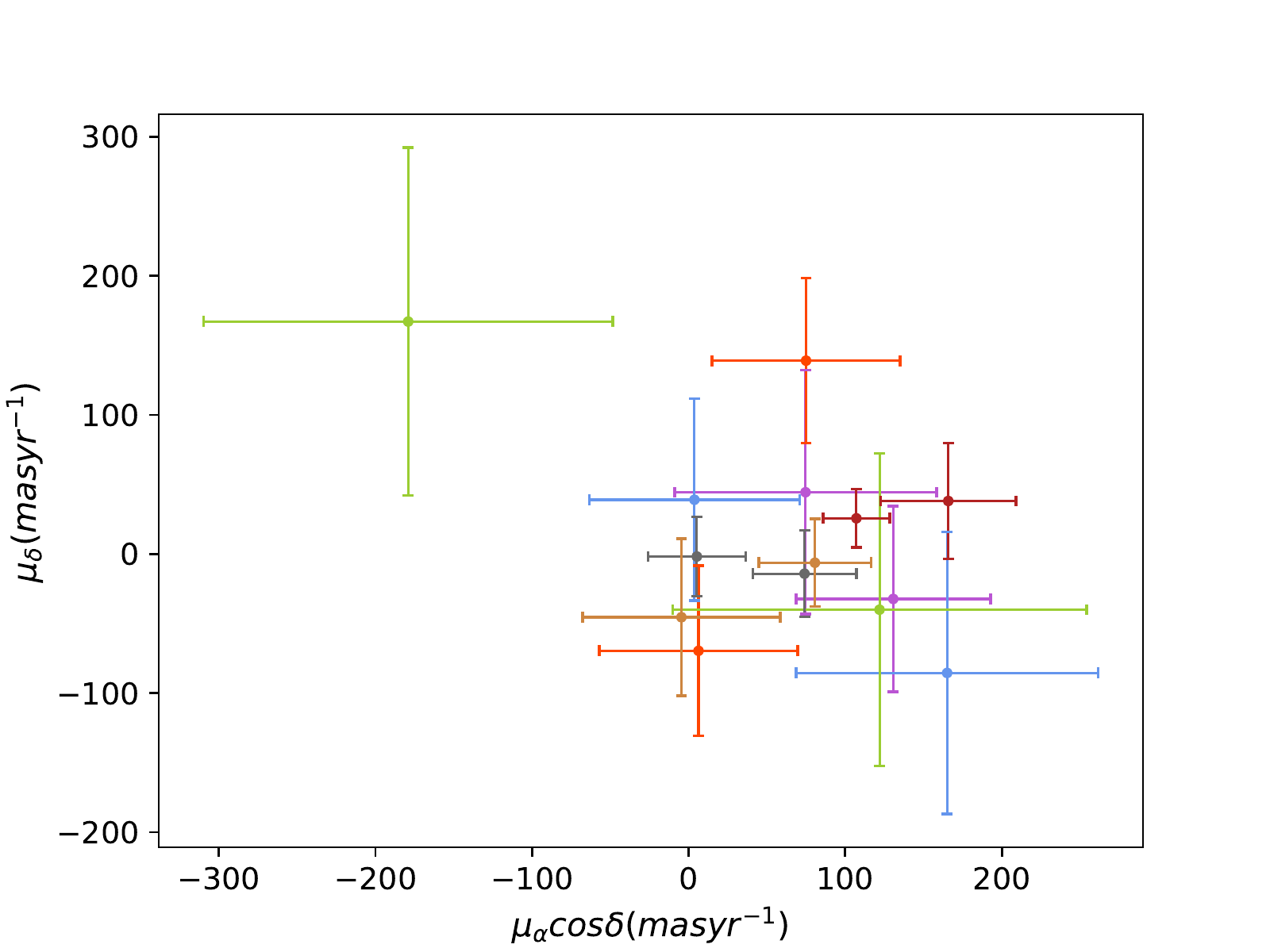}
\end{center}
\caption{Vector point diagram for the seven UCD pairs that satisfy the common distance and common proper motion criteria. Each pair is indicated by a different colour. The error bars correspond to an uncertainty of 1$\sigma$.}
\label{fig:ltlt_pm}
\end{figure}

We also used the CatWISE catalog to obtain the proper motion information for the wide binary involving two UCDs. One L0 member has proper motion from Gaia DR2. All nine pairs have proper motion measurements and seven remain in the sample after application of the proper motion filter. Figure \ref{fig:ltlt_pm} shows the vector point diagram for these seven pairs.
For more details regarding these systems visit \url{https://des.ncsa.illinois.edu/releases/other/y3-lt-widebinaries}.

\begin{figure*}
\begin{center}
   \includegraphics[width=0.6\linewidth]{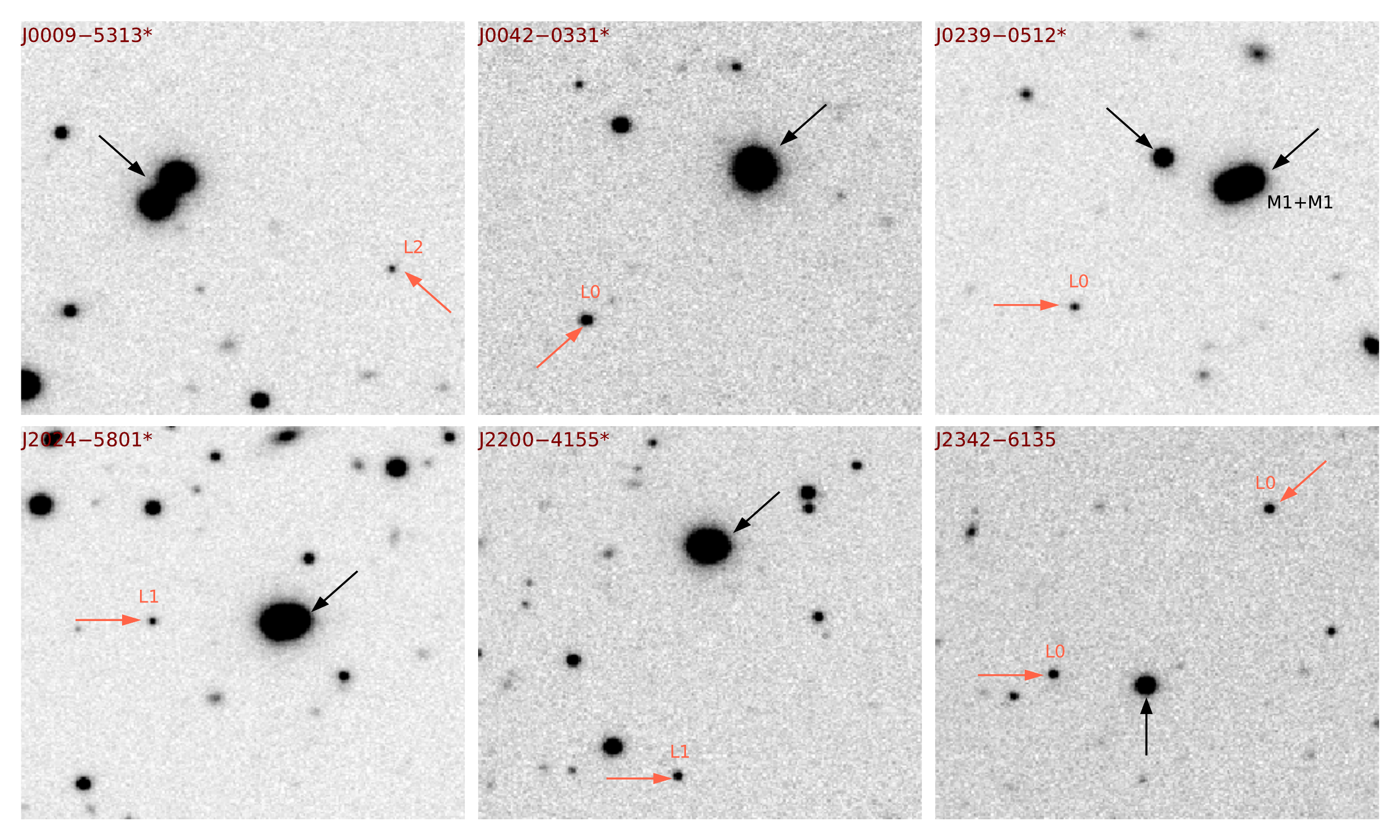}
\end{center}
\caption{$60\arcsec \times 60\arcsec$ $z$ band images of the multiple systems found. The black arrow indicates the stars, while the UCDs are identified by a red arrow followed by their spectral type. The upper right image corresponds to a quadruple candidate system. The double M1+M1 were previously identified by \citet{Dhital2015}. This quadruple system also contains a common distance between the L0 and a star, indicated by the arrows. The remaining images correspond to candidates of triple systems. The lower right panel corresponds to two UCD companions to a star.}
\label{fig:moisac-multi}
\end{figure*}

\subsection{Multiple systems}
\label{sec:search_multiple}

In addition to our wide binary candidates presented in Section \ref{sec:search_gaia} and Section \ref{sec:search_desvhs}, we find several possible multiple systems: five triple and one quadruple system. All members of the candidate triple systems satisfy the common distance criterion when considered two by two. As for the quadruple candidate, in \citet{Dhital2015} the system is presented as an M1+M1 binary, but we identified two more members. In this case, the L0 member does not satisfy the common distance criterion with one of the M1 stars in the binary reported by \citet{Dhital2015} and it marginally satisfies this criterion with the other M1.

As for proper motions, all six systems have proper motion measurements for all members. We again use the CatWISE catalog to obtain the proper motion for the L dwarfs. Applying the common proper motion criteria, we discard the quadruple as a physical system. The M1+M1 binary does not have common proper motion with the other stellar member. The proper motion of the L0 is consistent with the brighter three sources, but has an uncertainty comparable to its value and therefore is not very informative. As for the triple systems, the four pairs within them all satisfy the common proper motion criteria presented in the beginning of Section \ref{sec:search}, except for J2024-5801, where the binary star has a difference in measured proper motions beyond $2 \sigma$. However, the expected motion caused by a physical pair orbiting at their separation is comparable to this observed difference. One of the triple systems, J2342-6135, is composed by two UCDs and a stellar member. Again, the very large uncertainty of the UCD proper motion prevent stronger conclusions about this system. The candidate multiple systems are shown in Figure \ref{fig:moisac-multi} and their main characteristics are described in Table \ref{tab:multi_sys}. For more details regarding the table content visit \url{https://des.ncsa.illinois.edu/releases/other/y3-lt-widebinaries}.  

For the multiple systems, the chance alignment probability requires estimating and combining the probabilities of random alignment of each of the three (in case of triples) or six (in case of quadruples) pairs involved in the system, as well as the chance alignments of higher orders up to that of the entire system altogether. As this would involve much larger simulations sets, we refrain from computing the chance alignment probabilities for these systems. However, the configurations of the quadruple system, with no clear hierarchy, and of the triple system with two UCDs, are both very uncommon. Combined with the previous discussion based on proper motions, this is a clear indication that these systems are not physical and that the algorithm based on common distances is leading to the identification of physically unrelated systems of higher order, as in the case of the wide binaries.

\begin{figure*}
\begin{center}
    \includegraphics[width=0.65\linewidth]{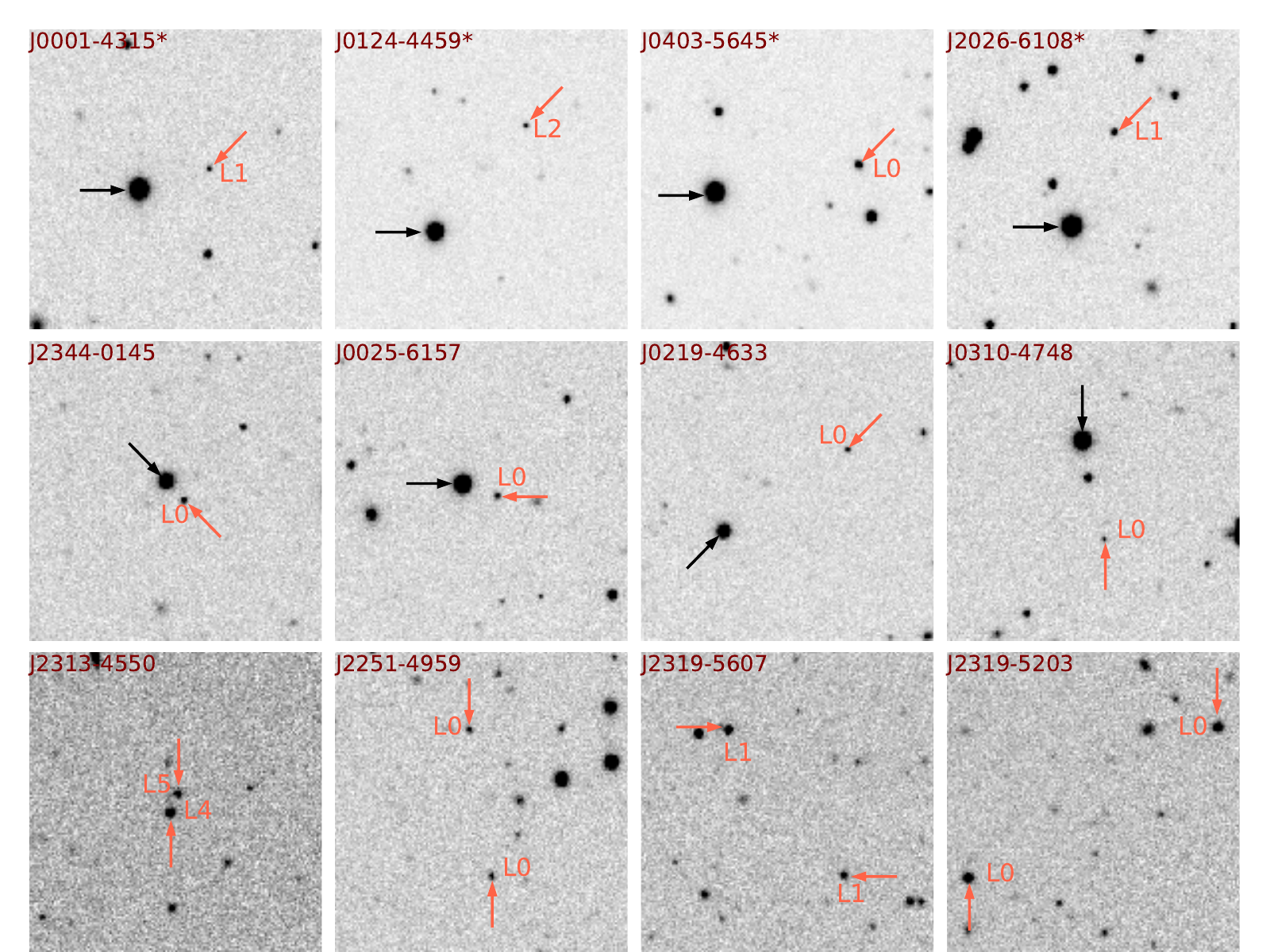}
\end{center}
\caption{$60\arcsec \times 60\arcsec$ z band images of selected binary candidates systems. In the first row, we present L dwarfs as companions of GaiaDR2-18 stars from \citet{Anders2019}. In the second row, the L dwarfs as companions of DES stars. In the last row, we present binary pairs composed by two UCDs. In all images the primary star is identified by an black arrow and the secondary by a red arrow followed by their spectral type.}
\label{fig:moisac}
\end{figure*}

\section{Discussion}
\label{sec:discuss}

For our 264 common distance pair candidates, we visually inspected the DES images. Figure \ref{fig:moisac} shows a sample of some selected binary candidates. The rows show pairs constituted by a UCD companion to a GaiaDR2-18 star, to a DES DR1 star and also systems made up by two UCDs, in this order. All of the images were taken from the DES Science Portal related to the DR1 public release images \footnote{\url{https://des.ncsa.illinois.edu/releases/dr1/dr1-access}}.

In Table \ref{tab:know_obj_literature} we present the known F/G/K/M+L or T wide systems already published in the literature that were spectroscopically confirmed and have an UCD as a companion. In Table \ref{tab:cmcp_literature} we present the common distance and/or common proper motion known F/G/K/M+L or T wide systems identified so far. Using this information, we searched for matches between our pairs and multiple system candidates presented in this work and the previously known pairs, but none of the 264 pairs and six multiples were identified among them. The main reason is that the majority of the known wide binaries with spectroscopic confirmation are in the northern hemisphere and/or have a projected separation < 600 AU and we are not able to resolve them. 

We also perform a search using the catalogue SLoWPoKES I and II presented in \citet{Dhital2011} and \citet{Dhital2015}, respectively, which contains low mass stars wide binaries identified using common distance and/or common proper motion. In this case, we were able to identify one M1+M1 common system as discussed in Section \ref{sec:search_multiple}.

\begin{figure}
\begin{center}
    \includegraphics[width=\linewidth]{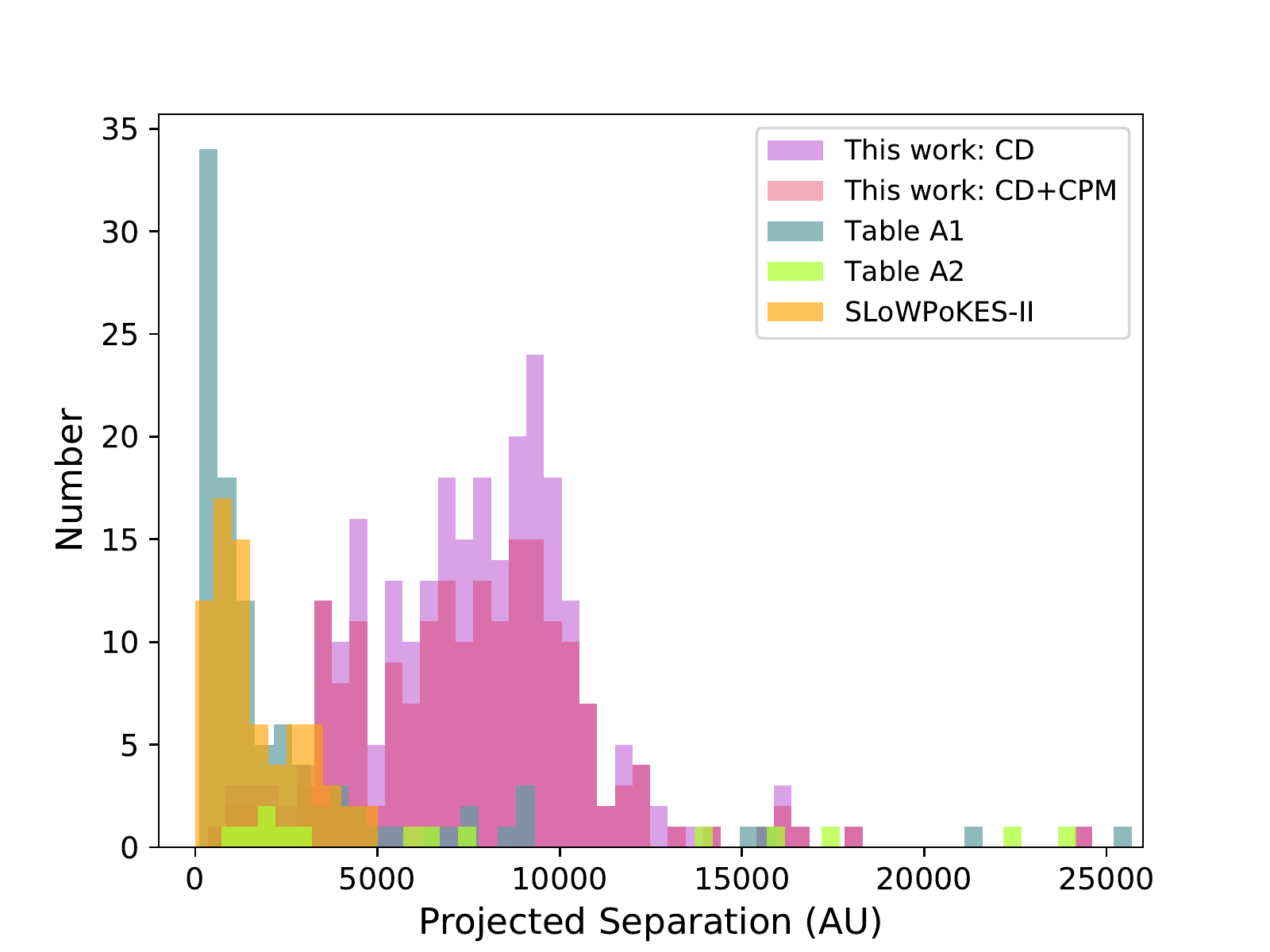}
\end{center}
\caption{Distribution of projected separations using four different samples, as indicated in the upper right corner. The CD and CPM labels mean common distance and common proper motion, respectively. Our wide binary sample is the most numerous and reaches larger projected separation than the previous known samples. Table \ref{tab:cmcp_literature} has unbound systems with very large projected separations. In order to better understand the distribution of separations, the figure only contains objects with a limiting of 26,000 AU in separation.}
\label{fig:projected_separation}
\end{figure}

Figure \ref{fig:projected_separation} shows the distributions of projected separations from our wide binary candidates sample, the 141 SLoWPoKES-II wide very low mass binaries, and from Tables \ref{tab:know_obj_literature} and \ref{tab:cmcp_literature}. Our sample was divided into binary systems that satisfy the common distance criterion alone and those that satisfy the common distance plus common proper motion criteria. The projected separations in our sample are those listed in the Tables \ref{tab:gaia_lt} and \ref{tab:ltlt} and were computed using the angular separation and the primary's distance. For this reason they may exceed the 10,000 AU limit originally set for the search radius, which was based on the lowest boundary in distance given the uncertainties. The distributions are all different from each other, reflecting selection biases. Spectroscopically confirmed systems containing UCDs are largely restricted to small separations compared to common distance and common proper motions pairs. Our samples, with and without the common proper motion criterion, also span larger separations than those from \citet{Dhital2015}. In fact, adding the proper motion constraint barely changes the shape of the distribution of projected separations, but clearly reduces the number of objects due to lack of proper motion data. Furthermore, as discussed in Section \ref{sec:search}, the very large uncertainties in the proper motions of most UCDs from CatWISE, indicate that the currently available proper motions are not an efficient diagnostic in this case. Therefore, in the subsequent analyses, we will adopt the common distance objects as our final sample. The abrupt drop in the number of systems with separations $> 10,000$ AU is due to our search radius limit.

Our sample is also the largest of those shown, given the larger photometric and astrometric samples it is derived from. As discussed previously by \citet{Dhital2015}, a large number of wide binary low-mass systems in the Galactic field could rule out the proposed formation scenario where very low mass objects are ejected from the protocluster due to dynamical interactions \citep{Reipurth2001, Bate2005}. Due to their low binding energy, they are unlikely to survive this dynamical process.

\begin{figure}
\begin{center}
    \includegraphics[width=\linewidth]{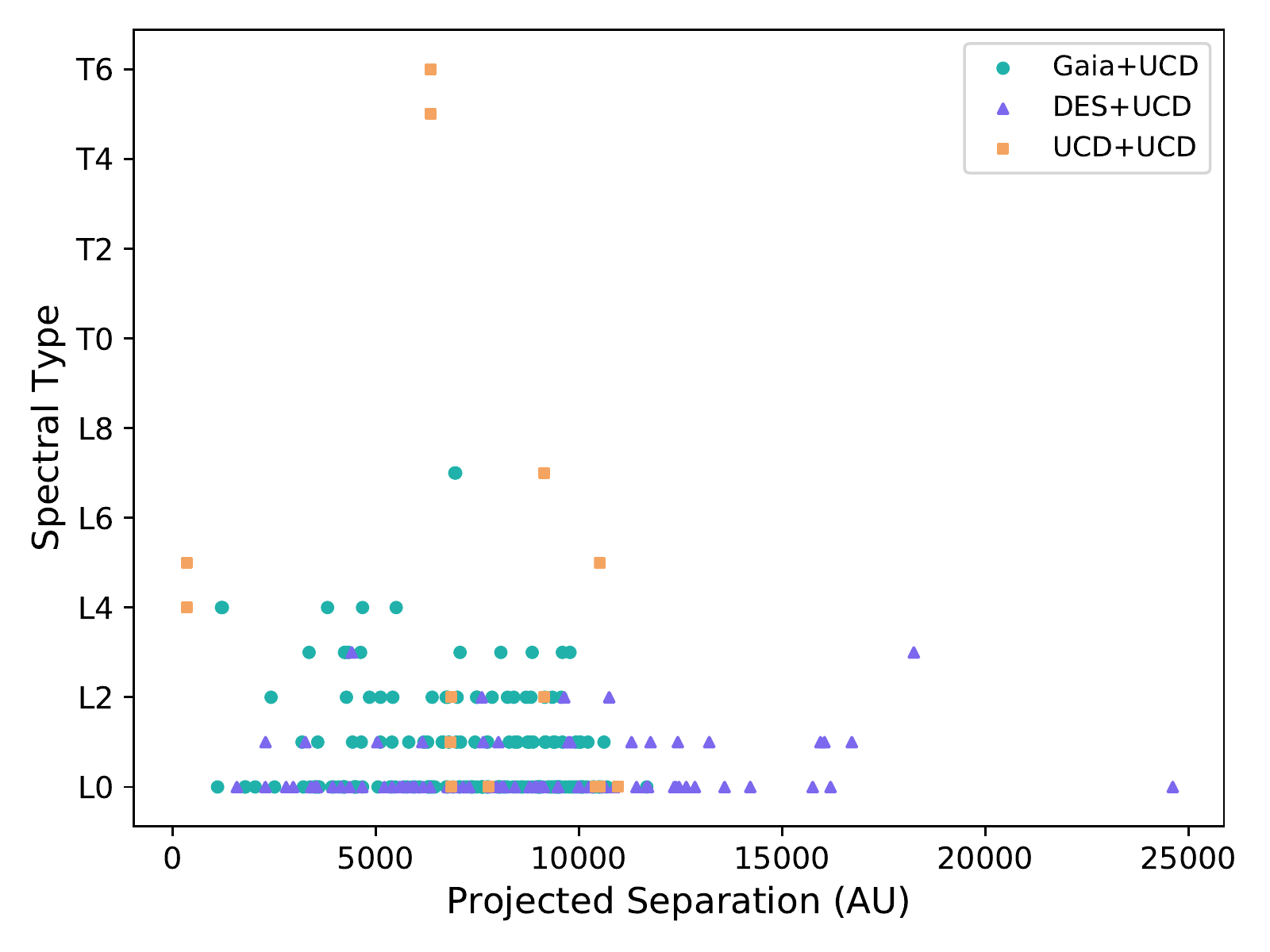}
\end{center}
\caption{Spectral type of the ultracool dwarfs plotted against the projected separation of the common distance pairs. The green dots and the purple triangles represent the companions of GaiaDR2-18 and DES stars, respectively. The orange boxes indicate the systems composed by two ultracool dwarfs.}
\label{fig:spt_separation}
\end{figure}

Figure \ref{fig:spt_separation} shows the spectral type of the UCDs versus the projected separation of the common distance pairs. Our sample of wide binary candidates contains 271 L dwarfs companions to stars with projected separations ranging from > 1,000 AU to 24,000 AU. We have nine wide systems made up by two UCDs that satisfy the common distance criterion and seven of them also satisfy the common proper motion criterion. If confirmed, these will be the widest systems (> 6,000 AU) involving two L/T dwarfs currently known. Only one candidate double T dwarf system was found, with a projected separation $\sim$ 6,000 AU. \citet{Deacon2014} pointed out the paucity of T dwarfs companions wider than 3,000 AU, which means that this system may be a rare find.

\begin{figure}
\begin{center}
    \includegraphics[width=\linewidth]{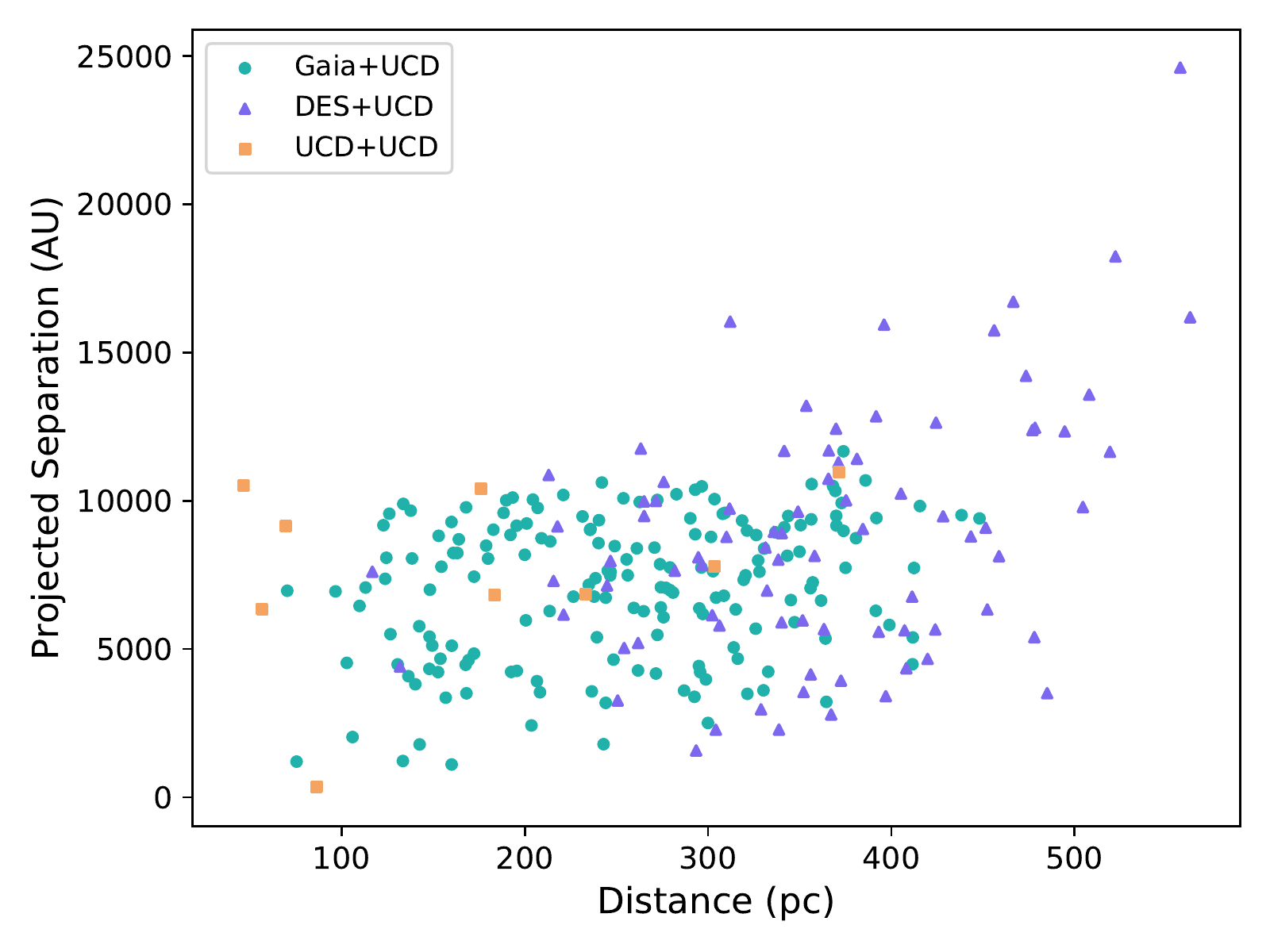}
\end{center}
\caption{Projected pair separation in AU plotted against distance for the 264 common distance binary candidates with $P_{a}$ < 5 \%. The colours and different symbols represent the three different samples presented previously, as indicated in the upper left corner. The zone of avoidance at small projected separations ($< 1$ AU) is caused by spatial resolution limits, while the scarcity of pairs with separations larger than 10,000 AU, specially for distances smaller than $\simeq 300$ pc, is due to the search method.}
\label{fig:dist_separation}
\end{figure}

Figure \ref{fig:dist_separation} shows the projected separations against distances for our common distance candidate wide binary sample. It is limited to $\sim$ 500 pc, making it the deepest sample of binaries involving UCDs. 82\% of our pair candidates concentrate at a distance < 400 pc and projected separation < 10,000 AU as shown in the Figure. This is in part due to the fact that the chance alignment probability tends to grow with the projected separation and the heliocentric distance of the primary.

Figure \ref{fig:dist_separation} also shows a lower limit in projected separation which is related to the typical angular resolution of the DES DR1 and Gaia DR2 images, especially the former, from which the binary sample is drawn. Pairs whose angular separation is of the order or lower than the DES seeing limit will be harder to resolve. At a distance of 480 pc, a 1.3 arcsec resolution limit will translate into a minimum separation of $\simeq 620$ AU, which is roughly what Figure \ref{fig:dist_separation} shows as a lower limit.

\begin{figure}
\begin{center}
    \includegraphics[width=\linewidth]{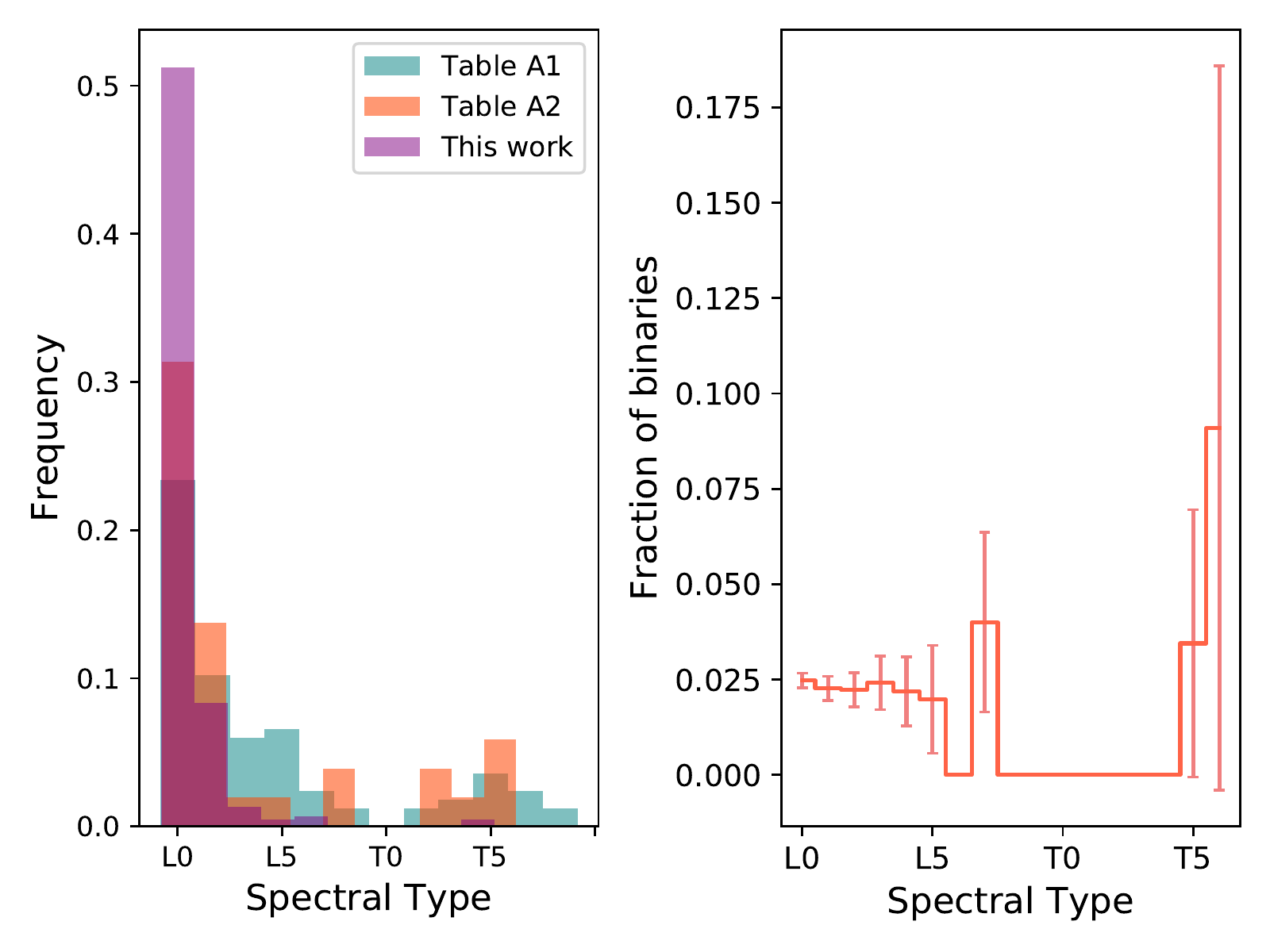}
\end{center}
\caption{The left panel shows the frequency distribution of UCDs in wide binary systems, considering our sample and the currently known systems. The right panel shows the observed fraction of wide binaries (in the separation range as shown in Figure \ref{fig:dist_separation}) as a function of spectral type. The error bars are Poissonian.}
\label{fig:hist_samples}
\end{figure}

Using the wide binary systems presented in Table \ref{tab:know_obj_literature}, Table \ref{tab:cmcp_literature} and our sample, we compared the frequency distribution of spectral types, as presented in left panel of Figure \ref{fig:hist_samples}. The L dwarfs dominate all samples. Our common distance sample is particularly biased towards early L types, as expected for the optical data on which our selection of UCDs and of primary stars is based. This sample represents a very significant leap compared to the number of such systems known. Even in a deep optical survey such as DES, we are still bound to detect mainly L types at $\sim$ 500 pc and this selection bias against later types clearly appears in the distributions. The right panel shows the fraction of candidate wide binaries (within the projected separation limits discussed earlier) as a function of spectral type. We observe that the typical wide binary fraction is 2-4\% over most of the spectral types, specially among L dwarfs, where we have better statistics. We also have added Poisson uncertainties to the binary fractions for each spectral type as shown in the Figure \ref{fig:hist_samples}.

As for the candidate triple systems, it is interesting to notice that four of them have a similar configuration, with a tight binary plus a detached third member as a UCD. Systems with a very similar configuration to our findings have been previously reported in the literature, as in \citet{Kirkpatrick2001, Gomes2013, Dupuy2018, Gauza2019}. Regarding the formation scenarios, this type of system is consistent with results of dynamical modelling of three-body interactions including UCDs \citep{Delgado2004, Bate2012}.

\section{Summary and Conclusions}
\label{sec:summary}

Using the Gaia DR2 and the combination of DES, VHS and AllWISE data along with a sample of UCD candidates from \citet{Carnero2019}, we identified 264 new wide binary candidates. The projected separations for the wide binary systems are spread within the $\sim$ 1,000-24,000 AU range. The upper limit in projected distance results from our search strategy, in which we avoided larger separations that are more likely to be affected by contaminants. The lower limit in separation stems from the typical resolution of the DES images on which the original UCD sample is based. A sample of six candidate multiple systems were also identified and the projected separations between the UCD dwarfs and the stellar members of these higher order systems range from $\sim$ 3,000-11,000 AU. 

Our candidates were selected based on common distance criteria and with a chance alignment probability criterion of $ P_{a} < 5\%$. We also used proper motions from Gaia DR2 and from the CatWISE Catalog as an attempt to refine the sample. We found proper motion measurements for about $90\%$ of the sources in the pairs and multiple systems, and 73\% of them also satisfy common proper motion criteria as discussed in Section \ref{sec:search}. But the proper motion data still have large uncertainties regarding the UCDs. Most of the systems with proper motions available, however, have proper motions within 2$\sigma$ of each other. 

We found 174 common distance candidate pairs with a primary from the Gaia DR2 catalogue limited to $G < 18$, for which distances are estimated from the \verb+StarHorse+ code by \citet{Anders2019}. We also found 81 common distance candidate pairs with a primary from the DES DR1 sample. These latter tend to be fainter and their \verb+StarHorse+ distances are based mostly on photometry, although some have Gaia DR2 parallax information as well. In addition, we found nine systems containing two ultracool dwarfs. Hence, we found in total 264 new wide binary candidates. This is the largest sample of candidate wide binary systems to date and is also the one that reaches the largest distances. These binary and multiple system candidates involving very low mass and substellar sources are crucial as possible benchmarks to evolutionary models close to or below the hydrogen burning limit, since properties such as metallicity and age, as well as masses, may be obtained for the primaries. The large number of wide binaries found in this work is inconsistent with the formation of very low mass stars and brown dwarfs from strong dynamical interactions leading to their ejection of star forming cores, since the binding energy involved is very low and would lead in most cases to the pair dissolution.

We also found six possible multiple systems, of which five are triples and one is a quadruple. The only potential quadruple system found is composed of an L0 dwarf associated to a star and to an M1+M1 double found previously by \citet{Dhital2015}, but the L0's distance is only marginally consistent with that of the M1+M1 double, while the third star has a proper motion that is inconsistent with that of the brighter pair. One of the five triples is composed by two L dwarfs associated with a DES star companion. The configuration of both the quadruple and of this triple is also very atypical of multiple systems, again making their physical reality unlikely. On the other hand, the other four triple systems show a similar configuration, with a tight pair and a detached third object. This is also commonly seen in other triple systems reported in the literature, and is a favoured configuration according to models of three-body encounters \citep{Delgado2004, Bate2012}.

\begin{table}
    \centering
    \caption{Summary of the common distance systems found. The systems with chance alignment probability >5\% are not included here. CD and CPM stand for common distance and common proper motion, respectively. The PM column indicated how many CD systems have proper motion measurements.}
    \label{tab:summary}
    \begin{tabular}{lcccccc}
        \hline \hline
        Type of system & & \multicolumn{3}{c}{Total} \\
        & & CD & PM & CD+CPM\\
        \hline
        \multirow{3}*{Binary} 
               & Gaia+UCD & 174 & 153 & 125 \\
               & DES+UCD & 81 & 74 & 61\\
               & UCD+UCD & 9 & 9 & 7\\
        Triple & & 5 & 5 & 4 \\
        Quadruple & & 1 & 1 & -\\
        \hline
    \end{tabular}
    
\end{table}

Table \ref{tab:summary} summarizes all the systems found in this work, regarding its type and the total number of systems, with and without proper motion data available. About 64\% of our ultracool dwarfs found in candidate binary and multiple systems are of the L0 spectral type. Still they make up only $\simeq 2\%$ of the total sample of L0 by \citet{Carnero2019}. The typical wide binary fraction for the binary candidates over all spectral types ranges from $2-4\%$ in the projected separation range covered by this work. The wide binary systems with UCDs as members presented here comprehend the largest catalogue to date. 

Given the measurements of the chance alignment probabilities above, we expect some physically unrelated systems to remain in our sample.
The systems here identified, therefore, must all be considered as binary or multiple system candidates, pending on kinematical and spectroscopic confirmation. Still, this catalogue constitutes a significant leap in the number of candidate wide separation systems containing UCDs and in the estimates of the wide binary fraction for UCDs. Evolutionary models predict that our sample dominated by early L sources should include young or intermediate age brown dwarfs, whose benchmarking may also be very useful to constrain models.

\section*{Acknowledgments}

Funding for the DES Projects has been provided by the U.S. Department of Energy, the U.S. National Science Foundation, the Ministry of Science and Education of Spain, 
the Science and Technology Facilities Council of the United Kingdom, the Higher Education Funding Council for England, the National Center for Supercomputing 
Applications at the University of Illinois at Urbana-Champaign, the Kavli Institute of Cosmological Physics at the University of Chicago, the Center for Cosmology and Astro-Particle Physics at the Ohio State University, the Mitchell Institute for Fundamental Physics and Astronomy at Texas A\&M University, Financiadora de Estudos e Projetos, 
Funda{\c c}{\~a}o Carlos Chagas Filho de Amparo {\`a} Pesquisa do Estado do Rio de Janeiro, Conselho Nacional de Desenvolvimento Cient{\'i}fico e Tecnol{\'o}gico and the Minist{\'e}rio da Ci{\^e}ncia, Tecnologia e Inova{\c c}{\~a}o, the Deutsche Forschungsgemeinschaft and the Collaborating Institutions in the Dark Energy Survey. 

The Collaborating Institutions are Argonne National Laboratory, the University of California at Santa Cruz, the University of Cambridge, Centro de Investigaciones Energ{\'e}ticas, 
Medioambientales y Tecnol{\'o}gicas-Madrid, the University of Chicago, University College London, the DES-Brazil Consortium, the University of Edinburgh, the Eidgen{\"o}ssische Technische Hochschule (ETH) Z{\"u}rich, Fermi National Accelerator Laboratory, the University of Illinois at Urbana-Champaign, the Institut de Ci{\`e}ncies de l'Espai (IEEC/CSIC), 
the Institut de F{\'i}sica d'Altes Energies, Lawrence Berkeley National Laboratory, the Ludwig-Maximilians Universit{\"a}t M{\"u}nchen and the associated Excellence Cluster Universe, the University of Michigan, the NSF's National Optical-Infrared Astronomy Research Laboratory, the University of Nottingham, The Ohio State University, the University of Pennsylvania, the University of Portsmouth, SLAC National Accelerator Laboratory, Stanford University, the University of Sussex, Texas A\&M University, and the OzDES Membership Consortium.

Based in part on observations at Cerro Tololo Inter-American Observatory, NSF's National Optical-Infrared Astronomy Research Laboratory, which is operated by the Association of Universities for Research in Astronomy (AURA) under a cooperative agreement with the National Science Foundation.

The DES data management system is supported by the National Science Foundation under Grant Numbers AST-1138766 and AST-1536171. The DES participants from Spanish institutions are partially supported by MINECO under grants AYA2015-71825, ESP2015-66861, FPA2015-68048, SEV-2016-0588, SEV-2016-0597, and MDM-2015-0509, some of which include ERDF funds from the European Union. IFAE is partially funded by the CERCA program of the Generalitat de Catalunya. Research leading to these results has received funding from the European Research
Council under the European Union's Seventh Framework Program (FP7/2007-2013) including ERC grant agreements 240672, 291329, and 306478. We  acknowledge support from the Australian Research Council Centre of Excellence for All-sky Astrophysics (CAASTRO), through project number CE110001020, and the Brazilian Instituto Nacional de Ci\^encia e Tecnologia (INCT) e-Universe (CNPq grant 465376/2014-2).

This manuscript has been authored by Fermi Research Alliance, LLC under Contract No. DE-AC02-07CH11359 with the U.S. Department of Energy, Office of Science, Office of High Energy Physics. The United States Government retains and the publisher, by accepting the article for publication, acknowledges that the United States Government retains a non-exclusive, paid-up, irrevocable, world-wide license to publish or reproduce the published form of this manuscript, or allow others to do so, for United States Government purposes.

This publication makes use of data products from the Wide-field Infrared Survey Explorer, which is a joint project of the University of California, Los Angeles, and the Jet Propulsion Laboratory/California Institute of Technology, and NEOWISE, which is a project of the Jet Propulsion Laboratory/California Institute of Technology. WISE and NEOWISE are funded by the National Aeronautics and Space Administration. 

The analysis presented here is based on observations obtained as part of the VISTA Hemisphere Survey, ESO Programme, 179.A-2010 (PI: McMahon). 

This paper has gone through internal review by the DES collaboration.

ACR acknowledges financial support provided by the PAPDRJ CAPES/FAPERJ Fellowship and by ``Unidad de Excelencia Mar\'ia de Maeztu de CIEMAT - F\'isica de Part\'iculas (Proyecto MDM)".

\section{Data availability}

Data underlying this article are available in \url{https://des.ncsa.illinois.edu/releases/other/y3-lt-widebinaries}.

%\nocite{*}
%\bibliographystyle{mn2e}
%\bibliographystyle{apalike}
\bibliography{references}

\section{Affiliations}

$^{1}$ Instituto de F\'\i sica, UFRGS, Caixa Postal 15051, Porto Alegre, RS - 91501-970, Brazil \\
$^{2}$ Laborat\'orio Interinstitucional de e-Astronomia - LIneA, Rua Gal. Jos\'e Cristino 77, Rio de Janeiro, RJ - 20921-400, Brazil \\
$^{3}$ Centro de Investigaciones Energ\'eticas, Medioambientales y Tecnol\'ogicas (CIEMAT), Madrid, Spain \\
$^{4}$ Center for Astrophysics Research, University of Hertfordshire, Hatfield AL10 9AB, UK \\
$^{5}$ Fermi National Accelerator Laboratory, P. O. Box 500, Batavia, IL 60510, USA \\
$^{6}$ George P. and Cynthia Woods Mitchell Institute for Fundamental Physics and Astronomy, and Department of Physics and Astronomy, Texas A\&M University, College Station, TX 77843, USA \\
$^{7}$ LSST, 933 North Cherry Avenue, Tucson, AZ 85721, USA \\
$^{8}$ Physics Department, 2320 Chamberlin Hall, University of Wisconsin-Madison, 1150 University Avenue Madison, WI \\
$^{9}$ Center for Cosmology and Astro-Particle Physics, The Ohio State University, Columbus, OH 43210, USA \\
$^{10}$ Department of Astronomy, The Ohio State University, Columbus, OH 43210, USA \\
$^{11}$ Department of Astrophysical Sciences, Princeton University, Peyton Hall, Princeton, NJ 08544, USA \\
$^{12}$ Kavli Institute for Cosmological Physics, University of Chicago, Chicago, IL 60637, USA \\
$^{13}$ Observatories of the Carnegie Institution for Science, 813 Santa Barbara St., Pasadena, CA 91101, USA \\
$^{14}$ Cerro Tololo Inter-American Observatory, NSF's National Optical-Infrared Astronomy Research Laboratory, Casilla 603, La Serena, Chile \\
$^{15}$ Departamento de F\'isica Matem\'atica, Instituto de F\'isica, Universidade de S\~ao Paulo, CP 66318, S\~ao Paulo, SP, 05314-970, Brazil \\
$^{16}$ Instituto de Fisica Teorica UAM/CSIC, Universidad Autonoma de Madrid, 28049 Madrid, Spain \\
$^{17}$ CNRS, UMR 7095, Institut d'Astrophysique de Paris, F-75014, Paris, France \\
$^{18}$ Sorbonne Universit\'es, UPMC Univ Paris 06, UMR 7095, Institut d'Astrophysique de Paris, F-75014, Paris, France \\
$^{19}$ Department of Physics and Astronomy, Pevensey Building, University of Sussex, Brighton, BN1 9QH, UK \\
$^{20}$ Department of Physics \& Astronomy, University College London, Gower Street, London, WC1E 6BT, UK \\
$^{21}$ Department of Astronomy, University of Illinois at Urbana-Champaign, 1002 W. Green Street, Urbana, IL 61801, USA \\
$^{22}$ National Center for Supercomputing Applications, 1205 West Clark St., Urbana, IL 61801, USA \\
$^{23}$ Institut de F\'{\i}sica d'Altes Energies (IFAE), The Barcelona Institute of Science and Technology, Campus UAB, 08193 Bellaterra (Barcelona) Spain \\
$^{24}$ Observat\'orio Nacional, Rua Gal. Jos\'e Cristino 77, Rio de Janeiro, RJ - 20921-400, Brazil \\
$^{25}$ Department of Astronomy/Steward Observatory, University of Arizona, 933 North Cherry Avenue, Tucson, AZ 85721-0065, USA \\
$^{26}$ Jet Propulsion Laboratory, California Institute of Technology, 4800 Oak Grove Dr., Pasadena, CA 91109, USA \\
$^{27}$ Santa Cruz Institute for Particle Physics, Santa Cruz, CA 95064, USA \\
$^{28}$ Institut d'Estudis Espacials de Catalunya (IEEC), 08034 Barcelona, Spain \\
$^{29}$ Institute of Space Sciences (ICE, CSIC),  Campus UAB, Carrer de Can Magrans, s/n,  08193 Barcelona, Spain \\
$^{30}$ Department of Astronomy, University of Michigan, Ann Arbor, MI 48109, USA \\
$^{31}$ Department of Physics, University of Michigan, Ann Arbor, MI 48109, USA \\
$^{32}$ Department of Physics, Stanford University, 382 Via Pueblo Mall, Stanford, CA 94305, USA \\
$^{33}$ SLAC National Accelerator Laboratory, Menlo Park, CA 94025, USA \\
$^{34}$ School of Mathematics and Physics, University of Queensland,  Brisbane, QLD 4072, Australia \\
$^{35}$ Department of Physics, The Ohio State University, Columbus, OH 43210, USA \\
$^{36}$ Center for Astrophysics $\vert$ Harvard \& Smithsonian, 60 Garden Street, Cambridge, MA 02138, USA \\
$^{37}$ Australian Astronomical Optics, Macquarie University, North Ryde, NSW 2113, Australia \\
$^{38}$ Lowell Observatory, 1400 Mars Hill Rd, Flagstaff, AZ 86001, USA \\
$^{39}$ Department of Physics and Astronomy, University of Pennsylvania, Philadelphia, PA 19104, USA \\
$^{40}$ Instituci\'o Catalana de Recerca i Estudis Avan\c{c}ats, E-08010 Barcelona, Spain \\
$^{41}$ School of Physics and Astronomy, University of Southampton,  Southampton, SO17 1BJ, UK \\
$^{42}$ Computer Science and Mathematics Division, Oak Ridge National Laboratory, Oak Ridge, TN 37831 \\
$^{43}$ Institute of Cosmology and Gravitation, University of Portsmouth, Portsmouth, PO1 3FX, UK \\
$^{44}$ Max Planck Institute for Extraterrestrial Physics, Giessenbachstrasse, 85748 Garching, Germany \\
$^{45}$ Universit\"{a}ts-Sternwarte, Fakult\"{a}t f\"{u}r Physik, Ludwig-Maximilians Universit\"{a}t M\"{u}nchen, Scheinerstr. 1, 81679 M\"{u}nchen, Germany

\appendix

\onecolumn

\section{Tables from the literature}
\label{app:tables}

\begin{longtable}{lllllllll}
\caption{Known systems which contain a L or T dwarf as a secondary, all are spectroscopically confirmed. All the systems presented here have projected separation > 100 AU. This table was based on Table 12 from \citet{Deacon2014}.} \label{tab:know_obj_literature}\\
\hline \hline
Object Name & \multicolumn{2}{c}{Separation} & Distance & \multicolumn{2}{c}{Spectral Type} & \multicolumn{1}{c}{Mass}  & \multicolumn{1}{c}{References} \\
\multicolumn{1}{l}{} & \multicolumn{1}{c}{$\arcsec$} & \multicolumn{1}{c}{AU} & \multicolumn{1}{c}{pc} & \multicolumn{1}{c}{Companion} & \multicolumn{1}{c}{Primary} & \multicolumn{1}{c}{$M_{\odot}$} \\
\hline
\endfirsthead
\multicolumn{8}{l}%
{{\bfseries \tablename\ \thetable{} -- continued from previous page}} \\
\hline \hline
Object Name  & \multicolumn{2}{c}{Separation} & Distance & \multicolumn{2}{c}{Spectral Type} & \multicolumn{1}{c}{Mass} &  \multicolumn{1}{c}{References} \\
\multicolumn{1}{l}{} & \multicolumn{1}{c}{$\arcsec$} & \multicolumn{1}{c}{AU} & \multicolumn{1}{c}{pc} &  \multicolumn{1}{c}{Companion} & \multicolumn{1}{c}{Primary} & \multicolumn{1}{c}{$M_{\odot}$}  \\
\hline 
\endhead
\endfoot
\endlastfoot
HD65216BC                 & 7.0      & 253     & 36.1     & M7+L2    & G5       & 0.08        & 1         \\ 
LP213-68Bab               & 14.0     & 230     & 16.4     & M8+L0    & M6.5     & 0.068-0.090 & 14,15     \\ 
BD+131727B                & 10.5     & 380     & 36.1     & M8+L0.5  & K5       & -           & 13        \\ 
HD221356BC                & 452.0    & 11900   & 26.3     & M8+L3    & F8       & 0.072       & 27        \\ 
HD221356D                 & 12.13    & 2050    & 169.0    & L1       & F8+M8+L3 & 0.073-0.085 & 32        \\ 
DENISJ0551-4434B          & 2.2      & 220     & 100.0    & L0       & M8.5     & 0.06        & 5         \\ 
Denis-PJ1347-7610B        & 16.8     & 418     & 24.8     & L0       & M0       & -           & 6         \\ 
HD89744B                  & 63.0     & 2460    & 39.0     & L0       & F7       & 0.077-0.080 & 7         \\ 
NLTT2274B                 & 23.0     & 483     & 21.0     & L0       & M4       & 0.081-0.083 & 8         \\ 
LP312-49B                 & 15.4     & 801     & 52.0     & L0       & M4       & -           & 9         \\ 
SDSSJ130432.93+090713.7B  & 7.6      & 374     & 49.2     & L0       & M4.5     & -           & 9         \\ 
SDSSJ163814.32+321133.5B  & 46.0     & 2420    & 52.6     & L0       & M4       & -           & 9         \\ 
1RXSJ235133.3+312720B     & 2.4      & 120     & 49.9     & L0       & M2       & 0.026-0.038 & 10        \\ 
2MASS12593933+0651255     & 23.86    & 1110    & 46.5     & L0       & M8       & 0.21        & 11        \\ 
2MASS09411195+3315060     & 7.44     & 244     & 32.7     & L0       & M5       & 0.23        & 11        \\ 
HIP2397B                  & 117.1    & 3970    & 33.9     & L0.5     & K5       & -           & 12        \\ 
HD253662B                 & 20.1     & 1252    & 62.2     & L0.5     & G8       & -           & 12        \\ 
2M0858+2710               & 15.6     & 780     & 50.0     & L0       & M4       & 0.074-0.081 & 28        \\ 
2M1021+3704               & 22.2     & 3000    & 135.     & L0       & M4       & 0.071-0.076 & 28        \\ 
2M1202+4204               & 7.3      & 310     & 42.4     & L0       & M6       & 0.074-0.081 & 28        \\ 
2M0005+0626               & 6.1      & 400     & 65.5     & L0       & M4.5     & 0.079-0.085 & 28        \\ 
2M1222+3643               & 20.7     & 1635    & 78.9     & L0       & M3       & 0.074-0.081 & 28        \\ 
GaiaJ0452-36A             & 115.3    & 15828   & 137.2    & L0       & M1       & 0.084-0.086 & 29        \\ 
2MASS0719-50              & 58.7     & 1609    & 27.4     & L0       & M3.5     & -           & 75        \\ 
2M0013-1816               & 118.1    & 7400    & 62.6     & L1       & M3       & 0.072-0.078 & 28        \\ 
2M1441+1856               & 51.1     & 4110    & 80.4     & L1       & M6       & 0.072-0.079 & 28        \\ 
HIP59933B                 & 38.1     & 2170    & 56.9     & L1       & F8       & -           & 12        \\ 
HIP63506B                 & 132.8    & 5640    & 42.4     & L1       & M0       & -           & 12        \\ 
HIP6407B                  & 44.9     & 2570    & 57.2     & L1+T3    & G5       & -           & 12        \\ 
GJ1048B                   & 11.9     & 250     & 21.0     & L1       & K2       & 0.055-0.075 & 16        \\ 
ABPicB                    & 5.5      & 275     & 50.0     & L1       & K2       & 0.01        & 17        \\ 
G124-62Bab                & 44.0     & 1496    & 34.0     & L1+L1    & dM4.5e   & 0.054-0.082 & 18        \\ 
HD16270                   & 11.9     & 254     & 21.3     & L1       & K3.5     & -           & 2,16,4    \\ 
GQLupB                    & 0.7      & 103     & 147.1    & L1       & K7       & 0.010-0.020 & 19        \\ 
ROX42Bb                   & 1.8      & 140     & 77.7     & L1       & M1       & 0.006-0.014 & 20,21     \\ 
LSPMJ0241+2553B           & 31.2     & 2153    & 69.0     & L1       & WD       & -           & 12        \\ 
HIP112422B                & 16.0     & 1040    & 65.0     & L1.5     & K2       & -           & 12        \\ 
LSPMJ0632+5053B           & 47.4     & 4499    & 94.9     & L1.5     & G2       & -           & 12        \\ 
PMI13518+4157B            & 21.6     & 613     & 28.3     & L1.5     & M2.5     & -           & 12        \\ 
NLTT44368B                & 90.2     & 7760    & 86.0     & L1.5     & M3       & -           & 12        \\ 
PMI22118-1005B            & 204.5    & 8892    & 43.4     & L1.5     & M2       & -           & 12        \\ 
HIP11161                  & 47.7     & 3300    & 69.1     & L1.5     & F5       & -           & 12        \\ 
$\eta$TelB                & 4.20     & 190     & -        & L1       & A0V      & 0.04        & 13       \\
$\beta$Cir                & 217.8    & 6656    & 30.5     & L1       & A3V      & 0.056       & 22        \\ 
HD164507AB                & 25.1     & 1136    & 45.2     & L1       & G5       & -           & 76        \\ 
V478Lyr                   & 17.05    & 462     & 27.0     & L1       & G8       & -           & 76        \\ 
2M0122+0331               & 44.8     & 2222    & 49.5     & L2       & G5       & 0.071-0.076 & 28        \\ 
NLTT1011B                 & 58.5     & 3990    & 68.2     & L2       & K7       & -           & 12        \\ 
G255-34B                  & 38.3     & 1364    & 35.6     & L2       & K8       & -           & 23        \\ 
2MASSJ05254550-7425263B   & 44.0     & 2000    & 45.4     & L2       & M3       & 0.06-0.075  & 24        \\ 
G196-3B                   & 16.2     & 300     & 18.5     & L2       & M2.5     & 0.015-0.04  & 25        \\ 
Gl618.1B                  & 35.0     & 1090    & 31.1     & L2.5     & M0       & 0.06-0.079  & 7         \\ 
HD106906b                 & 7.1      & 650     & 91.5     & L2.5     & F5       & 0.003-0.007 & 26        \\ 
HIP73169                  & 29.0     & 796     & 27.4     & L2.5     & M0       & -           & 12        \\ 
2MASSJ0249-0557AB         & 39.9     & 1950    & 48.8     & L2       & M6       & 0.010-0.012 & 39        \\ 
CD-288692                 & 50.91    & 2026    & 39.7     & L2       & K5       & -           & 76        \\ 
2MASSJ1839+4424           & 21.89    & 811     & 37.0     & L2       & M9       & -           & 76        \\ 
2MASSJ0139+8110AB         & 23.0     & 959     & 41.6     & L2       & L1       & -           & 76        \\
2MASSJ2325+4608AB         & 7.24     & 378     & 52.2     & L2       & M8       & -           & 76        \\
G63-33B                   & 66.0     & 2010    & 30.4     & L3       & K2       & 0.079-0.081 & 8         \\ 
G73-26B                   & 73.0     & 2774    & 38.0     & L3       & M2       & 0.079-0.081 & 8,9       \\ 
2MASSJ2126-8140           & 217.0    & 6900    & 31.7     & L3       & M2       & 0.014-0.011 & 49        \\ 
2MASSJ22501512+2325342    & 8.9      & 518     & 58.2     & L3       & M3       & -           & 50        \\ 
$\eta$CancriB             & 164.0    & 15020   & 91.5     & L3.5     & K3III    & 0.063-0.082 & 9         \\ 
NLTT27966                 & 15.9     & 630     & 39.6     & L4       & M5       & -           & 12        \\ 
LSPMJ1336+2541            & 121.7    & 8793    & 72.2     & L4       & M3       & -           & 12        \\ 
NLTT26746B                & 18.0     & 661     & 36.7     & L4       & M4       & -           & 12        \\ 
PMI13410+0542B            & 9.4      & 484     & 51.4     & L4       & M1       & -           & 12        \\ 
G171-58B                  & 218.0    & 9200    & 42.2     & L4+L4    & F8       & 0.045-0.083 & 8         \\ 
G200-28B                  & 570.0    & 25700   & 45.0     & L4       & G5       & 0.077-0.078 & 8         \\ 
LHS5166B                  & 8.43     & 160     & 18.9     & L4       & dM4.5    & 0.055-0.075 & 18        \\ 
1RXSJ1609-2105b           & 2.2      & 330     & 150.0    & L4       & M0       & 0.009-0.016 & 33        \\ 
2MASSJ0219–3925           & 3.96     & 156     & 39.3     & L4       & M6       & -           & 78        \\ 
2M1259+1001               & 7.65     & 345     & 45.0     & L4.5     & M5       & 0.057-0.074 & 28        \\ 
GJ1001Bc                  & 18.6     & 180     & 9.6      & L4.5+L4.5& M4       & 0.060-0.075 & 29,34,35  \\ 
Gl417Bab                  & 90.0     & 2000    & 22.2     & L4.5+L6  & G0+G0    & 0.02-0.05   & 29,36     \\ 
HIP26653                  & 27.0     & 753     & 27.8     & L5       & G5       & -           & 12        \\ 
2M1115+1607               & 18.1     & 660     & 36.4     & L5       & M4       & 0.056-0.073 & 28        \\ 
G203-50B                  & 6.4      & 135     & 21.0     & L5.0     & M4.5     & 0.051-0.074 & 37        \\ 
GJ499C                    & 516.0    & 9708    & 18.8     & L5       & K5+M4    & -           & 23        \\ 
G259-20B                  & 30.0     & 650     & 21.6     & L5       & M2.5     & -           & 38        \\ 
HD196180                  & 13.51    & 907     & 67.1     & L5       & A3V      & -           & 40        \\ 
HIP85365B                 & 294.1    & 8850    & 30.0     & L5.5     & F3       & -           & 12        \\ 
NLTT55219B                & 9.7      & 432     & 44.5     & L5.5     & M2       & -           & 12        \\ 
HIP9269B                  & 52.1     & 1300    & 24.9     & L6       & G5       & -           & 12        \\ 
NLTT31450B                & 12.3     & 487     & 39.5     & L6       & M4       & -           & 12        \\ 
LP261-75B                 & 13.0     & 450     & 34.6     & L6       & M4.5     & 0.019-0.025 & 41        \\ 
2MASSJ01303563-4445411B   & 3.28     & 130     & 39.6     & L6       & M9       & 0.032-0.076 & 42        \\ 
NLTT20346                 & 248.0    & 7700    & 31.0     & L7+L6.5  & M5+M6    & 0.070       & 47        \\ 
VHS1256-1257              & 8.06     & 102     & 12.6     & L7       & M7.5     & 0.010       & 43        \\ 
HD203030B                 & 11.0     & 487     & 44.2     & L7.5     & G8       & 0.012-0.031 & 44        \\ 
NLTT730                   & 233.6    & 5070    & 21.7     & L7.5     & M4       & -           & 12        \\ 
Gl337CD                   & 43.0     & 880     & 20.4     & L8+L8    & G8+K1    & 0.04-0.074  & 7,45      \\ 
Gl584C                    & 194.0    & 3600    & 18.5     & L8       & G1       & 0.045-0.075 & 46        \\ 
PMI23492+3458             & 34.9     & 949    & 27.1     & L9       & M2       & -           & 12        \\ 
HD46588B                  & 79.2     & 1420   & 17.9     & L9       & F7       & 0.045-0.072 & 48        \\ 
NLTT51469C                & 82.27    & 3800   & 46.1     & L9       & M3+M6    & -           & 77        \\ 
$\epsilon$IndiBaBb        & 402.0    & 1460   & 3.6      & T1+T6    & K5       & 0.06-0.073  & 53,54     \\ 
2MASSJ111806.99-064007.8B & 7.7      & 650    & 84.4     & T2       & M4.5     & 0.06-0.07   & 55        \\ 
HNPegB                    & 43.0     & 795    & 18.4     & T2.5     & G0       & 0.012-0.030 & 56        \\ 
2MASSJ0213+3648ABC        & 16.4     & 360    & 21.9     & T3       & M4.5+M6.5& 0.068       & 51        \\ 
GUPscB                    & 41.97    & 2000   & 47.6     & T3.5     & M3       & 0.07-0.13   & 57        \\ 
HIP38939B                 & 88.0     & 1630   & 18.5     & T4.5     & K4       & 0.018-0.058 & 58        \\ 
LSPMJ1459+0851B           & 365.0    & 21500  & 58.9     & T4.5     & DA       & 0.064-0.075 & 59        \\ 
SDSSJ0006-0852AB          & 27.41    & 820    & 29.9     & T5       & M7+M8.5  & 0.056       & 52        \\ 
LHS2803B                  & 67.6     & 1400   & 20.7     & T5       & M4.5     & 0.068-0.081 & 24,60     \\ 
HD118865B                 & 148.0    & 9200   & 62.1     & T5       & F5       & -           & 61        \\ 
HIP63510C                 & 103.0    & 1200   & 11.6     & T6       & M0.5     & -           & 62        \\ 
HIP73786B                 & 63.8     & 1230   & 19.2     & T6       & K5       & -           & 62,63     \\ 
LHS302B                   & 265.0    & 4500   & 16.9     & T6       & M5       & -           & 64        \\ 
G204-39B                  & 198.0    & 2685   & 13.5     & T6.5     & M3       & 0.02-0.035  & 8         \\ 
Gl570D                    & 258.0    & 1500   & 5.8      & T7       & K4+M2+M3 & 0.03-0.07   & 65        \\ 
HD3651B                   & 43.0     & 480    & 11.1     & T7.5     & K0       & 0.018-0.058 & 56,66     \\ 
SDSSJ1416+30B             & 9.0      & 135    & 15.0     & T7.5     & L6       & 0.03-0.04   & 67,68,69  \\ 
LHS2907B                  & 156.0    & 2680   & 17.1     & T8       & G1       & 0.019-0.047 & 38,70     \\ 
LHS6176B                  & 52.0     & 1400   & 26.9     & T8       & M4       & -           & 38,61     \\ 
Wolf1130B                 & 188.5    & 3000   & 15.9     & T8       & sdM1.5+DA& 0.020-0.050 & 71        \\ 
Ross458C                  & 102.0    & 1162   & 11.3     & T8.5     & M0.5+M7  & 0.005-0.0014& 72        \\ 
$\xi$UMaE                 & 510.0    & 4100   & 8.0      & T8.5     & F9+G0    & 0.014-0.038 & 61        \\ 
Wolf940B                  & 32.0     & 400    & 12.5     & T8.5     & M4       & 0.02-0.032  & 73        \\ 
WD0806-661                & 130.0    & 2500   & 19.2     & >Y0      & DQ       & 0.03-0.10   & 74        \\ 
\hline

\caption*{\textbf{References:} (1) \citet{Mugrauer}; (2) \citet{anderson2012}; (3) \citet{Forveille2004}; (4) \citet{Dup12}; (5) \citet{Billeres2005}; (6) \citet{PhanBao2008}; (7) \citet{wilson2001}; (8) \citet{Faherty2010}; (9) \citet{Zhang2010}; (10) \citet{Bowler2012}; (11) \citet{GalvezOrtiz2017}; (12) \citet{Deacon2014}; (13) \citet{Cruz2007}; (14) \citet{Gizis2000}; (15) \citet{Close2003}; (16) \citet{Gizis2001}; (17) \citet{Chauvin2005}; (18) \citet{Seifahrt2005}; (19) \citet{Neuhauser2005}; (20) \citet{Kraus2014}; (21) \citet{Currie2014}; (22) \citet{Smith2015}; (23) \citet{Gomes2013}; (24) \citet{Muzic2012}; (25) \citet{Rebolo1998}; (26) \citet{Bailey2014}; (27) \citet{Caballero2007}; (28) \citet{Baron2015}; (29) \citet{Zhang2019}; (30) \citet{Casagrande2011}; (31) \citet{Metchev2004}; (32) \citet{Caballero2007}; (33) \citet{Lafreniere2008}; (34) \citet{Golimowski2004}; (35) \citet{Martin1999}; (36) \citet{bouy2003}; (37) \citet{Radigan2008}; (38) \citet{Luhman2012}; (39) \citet{Dupuy2018}; (40) \citet{derosa2015}; (41) \citet{Reid2006}; (42) \citet{Dhital2011}; (43) \citet{Gauza2015}; (44) \citet{Metchev2006}; (45) \citet{Burgasser2005}; (46) \citet{Kirkpatrick2000}; (47) \citet{Faherty2011}; (48) \citet{Loutrel2011}; (49) \citet{Deacon2016}; (50) \citet{Desrochers2018}; (51) \citet{Deacon2017}; (52) \citet{Burgasser2012}; (53) \citet{Scholz2003}; (54) \citet{McCaughrean2004}; (55) \citet{Reyle2013}; (56) \citet{Luhman2007}; (57) \citet{Naud2014}; (58) \citet{Deacon2012a}; (59) \citet{Day-Jones2011}; (60) \citet{Deacon2012b}; (61) \citet{Bur13}; (62) \citet{Scholz2010a}; (63) \citet{Murray2011}; (64) \citet{Kirkpatrick2011}; (65) \citet{Burgasser2000}; (66) \citet{Mugrauer2006}; (67) \citet{Scholz2010b} ; (68) \citet{Burningham2010}; (69) \citet{Bowler2009}; (70) \citet{Pinfield2012}; (71) \citet{Mace2013}; (72) \citet{Goldman2010}; (73) \citet{Burningham2009}; (74) \citet{Luhman2011}; (75) \citet{Andrei2011}; (76) \citet{Marocco2020}; (77) \citet{Gauza2019}; (78) \citet{Artigau2015}. }
\end{longtable}

\newpage

\begin{table}
\begin{center}
\caption{The common distance and common proper motion wide systems identified in the literature.} \label{tab:cmcp_literature}
\begin{tabular*}{\textwidth}{l@{\extracolsep{\fill}}llllllll}
%\begin{tabular*}{\textwidth}{lcccccccc}
\hline \hline
Object Name & \multicolumn{2}{c}{Separation} & Distance & Sp. Type & Sp. Type & $\mu_{\alpha} \cos \delta$ & $\mu_{\delta}$ & References\\
ID & $\arcsec$ & AU & pc & Companion & Primary & (mas $yr^-1$) & (mas $yr^-1$) & \\
\hline
2MASSJ0223-5815       	   & 816    & 400000 & 49 $\pm$10   & L0    & M5	& 134.0$\pm$10       & 5.0$\pm$19         & 1 \\
2MASSJ1214+3721       	   & 1866   & 153000 & 82 $\pm$17   & L0    & -	    & -122.6 $\pm$10.6   & 82.0 $\pm$17       & 1\\
2MASSJ0939+3412	           & 2516   & 156000 & 62 $\pm$10   & L0    & -     & -107.1 $\pm$10.4   & -64.3 $\pm$12.6    & 1 \\      
ULASJ0255+0532		       & 207    & 29000  & 140 $\pm$26  & L0    & F5    & 28 $\pm$30         & 40 $\pm$30         & 2\\      
ULASJ0900+2930		       & 81     & 16000  & 197 $\pm$37  & L0    & M3.5  & -13 $\pm$10        & -27.8 $\pm$8.8     & 2\\      
ULASJ1222+1407		       & 96     & 6700   & 70 $\pm$13   & L0    & M4    & -74 $\pm$20        & -34 $\pm$20        & 2\\      
2MASSJ09175035+2944455     & 1684.7 & 67388  & 40.0         & L0    & F5    &-47.54 $\pm$2.506   & -65.776 $\pm$1.844 & 5 \\
2MASSJ0626+0029            & 3761   & 252000 & 67 $\pm$14   & L0.5  & -     & 84 $\pm$15         & -92 $\pm$15        & 1\\  
2MASSJ1632+3505$^1$        & 57     & 2000   & 37 $\pm$8    & L0.5  & K0    & 91.6 $\pm$9.7      & -65.3 $\pm$11.9    & 1 \\  
2MASSJ17073334+4301304     & 917.2  & 23847  & 26 $\pm$2    & L0.5  & -     & -210.6 $\pm$8.9    & -47.2 $\pm$7.2     & 3 \\      
2MASSJ16325610+3505076     & 57.0   & 1938   & 34.9927      & L0.7  & K0    & 89.153  $\pm$0.51  & -60.527 $\pm$0.615 & 5 \\
2MASSJ2037-4216       	   & 5294   & 270000 & 51 $\pm$10   & L1    & -     & 229 $\pm$10        & -391 $\pm$10       & 1 \\      
2MASSJ0518461-275645       & 1007.2 & 57399  & 57.9079      & L1.0  & -     & 32.194  $\pm$1.299 & -4.943  $\pm$1.447 & 5 \\
SDSSJ124514.95+120112.0    & 96.4   & 5948   & 61.7         & L1    & DA    & -10.582 $\pm$4.067 & -53.728 $\pm$2.44  & 5 \\
G151-59                    & 46     & 3100   & 118          & L1    & K0    & 179 $\pm$9         & 158 $\pm$10        & 6 \\
2MASSJ14493646+0533379     & 246    & 33702  & 137          & L1    & -     & -107 $\pm$10       & -135 $\pm$10       & 6 \\
2MASSJ02235464-5815067     & 1532.6 & 62749  & 40.943       & L1.5  & M3+M9 & 104.21  $\pm$1.085 & -17.379 $\pm$0.918 & 5 \\
ULASJ1330+0914$^2$         & 409    & 61000  & 149 $\pm$30  & L2    & G5    & -83 $\pm$37        & 10 $\pm$37         & 2\\   
WISEAJ134824.42-422744.9   & 410.1  & 13940  & 34 $\pm$2    & L2    & -     & -144.3 $\pm$ 6.6   & -77.1 $\pm$6.5     & 3 \\
2MASSJ01415823-4633574     & 2377.2 & 86641  & 36.4465      & L2    & -     & 115.673 $\pm$0.7   & -46.609 $\pm$0.665 & 5 \\
2MASSJ08430796+3141297     & 819.5  & 38926  & 47.5         & L2.5  & -     &-52.293  $\pm$3.438 & -43.35  $\pm$2.189 & 5 \\
2MASSJ23225299-6151275     & 16.6   & 714    & 43.0283      & L2.5  & M5    & 80.092  $\pm$1.447 & -81.969 $\pm$1.621 & 5 \\
2MASSJ21265040-8140293     & 217.5  & 7436   & 34.1924      & L3    & M1.0  & 56.511  $\pm$1.656 & -115.369$\pm$2.441 & 5 \\
SDSSJ095932.74+452330.5    & 846.7  & 32175  & 38 $\pm$6    & L3/L4 & M4.5  & -97.1 $\pm$5.2     & -144.5 $\pm$9.4    & 3 \\ 
2MASSJ00283943+1501418     & 917.2  & 36688  & 40 $\pm$3    & L4.5  & -     & 199.3 $\pm$12.8    & -34.5 $\pm$11.7    & 3 \\ 
2MASSJ23512200+3010540     & 934.9  & 22416  & 24 $\pm$3    & L5pec & K5    & 251.7 $\pm$8.5     & 4.3 $\pm$7.1       & 3 \\ 
2MASSJ0230-0225        	   & 5370   & 145000 & 27 $\pm$6    & L8    & K1    & 329 $\pm$16.8      & 51.3 $\pm$14.9     & 1 \\ 
WISEAJ104335.09+121312.0   & 1039.6 & 17673  & 17 $\pm$8    & L9    & -     & 10.5 $\pm$8.4      & -245.2 $\pm$9.1    & 3 \\ 
PSOJ330.3214+32.3686       & 77.1   & 2313   & 20.1$\pm$2.1 & T2.5  & M1    & 105$\pm$8          & 65$\pm$9           & 4\\  
PSOJ334.1193+19.8800       & 52.2   & 1566   & 30.7$\pm$3.2 & T3    & M4    & 120$\pm$8          & -72$\pm$99         & 4 \\
2MASSJ1244+1232        	   & 6217   & 286000 & 46 $\pm$8    & T4    & -     & -104.8 $\pm$8.6    & 4.5 $\pm$7.3       & 1 \\  
2MASSJ0758+2225            & 4758   & 157000 & 33 $\pm$8    & T6.5  & -     & -105 $\pm$8        & -62.8 $\pm$8.2     & 1 \\  
2MASSJ1150+0949            & 1283   & 77000  & 60 $\pm$27   & T6.5  & -     & -107.6 $\pm$17.1   & -31.9 $\pm$4.5     & 1 \\  
2MASSJ0915+0531            & 5394   & 178000 & 33 $\pm$6    & T7    & G+G   & -95 $\pm$5.5       & -57.7 $\pm$4.4     & 1 \\   
\hline
\end{tabular*}
\end{center}
\begin{tablenotes}
     \item \textbf{References:} {(1) \citet{smart2017}; (2) \citet{marocco2017}; (3) \citet{Kirkpatrick2016};  (4) \citet{Best2015} ;  (5) \citet{Smart2019}; (6) \citet{Smith2014}.}
     \item \textbf{$^1$} This is the only bound system in \citet{smart2017}.
     \item \textbf{$^2$} Classify as unlikely pair \citep{marocco2017}.
\end{tablenotes}
\end{table}

\label{lastpage}
\end{document}